\documentclass{jfm}

\usepackage{graphicx}
\usepackage{newtxtext}
\usepackage{newtxmath}
\usepackage{natbib}
\usepackage{xcolor}
\usepackage{hyperref}
\usepackage[normalem]{ulem}
\hypersetup{
    colorlinks = true,
    urlcolor   = blue,
    citecolor  = black,
}

\newcommand{\RomanNumeralCaps}[1]

\usepackage[outdir=./FIGURES]{epstopdf}
\usepackage{subfigure}
\newcommand{\vt}[1]{{\color{black}#1}} 
\usepackage{setspace}

\title{On linear stability of supersonic flow over a short compression corner at large ramp angles}

\author{Irmak T. ~Karpuzcu\aff{1}
 \corresp{\email{itk3@illinois.edu}},
 Vassilis Theofilis\aff{2},
 Deborah A. Levin\aff{1}
 }

 \affiliation{\aff{1}Aerospace Engineering, University of Illinois, Urbana-Champaign, IL 61801, USA
\aff{2}
Center for High-Speed Flight, Faculty of Aerospace Engineering, Technion - Israel Institute of Technology, Haifa 32000, Israel 
}
\begin{document}
\maketitle

\begin{abstract}
Linear stability of supersonic flow over a short compression corner with ramp angles 30$^\mathrm{o}$ and 42$^\mathrm{o}$ is investigated using Direct Simulation Monte Carlo (DSMC) and Linear Stability Theory (LST) at Mach number $3$, Reynolds number $11,200$ and low Knudsen number, $O(10^{-4})$. The two-dimensional base flows feature nonzero velocity slip and temperature jump and were found to be  steady and laminar at both ramp angles. Modal analysis revealed a previously unknown traveling three-dimensional global mode, the amplitude functions of which peak at the leading-edge and separation shocks and extend within the shear layer of the large laminar separation bubble formed on the short compression corner. This mode is linearly unstable at the higher ramp angle and stable at the lower one, \textcolor{black}{while the known stationary three-dimensional global mode which peaks at the laminar separation is also present in the spectrum,} but is (strongly) damped at both ramp angles. Three-dimensional DSMC simulations have fully confirmed the LST results, underlined (again) the significance of modeling the shock contribution in linear stability analyses of high-speed flow, and predicted the nonlinear evolution of the flow up to the generation of lambda vortices on the ramp, \vt{for the first time in the context of {\color{black}kinetic theory} simulations}.
\end{abstract}


\date{\today}



\section{Introduction}\label{sec:introduction}

Compression ramps are common features of vehicles traveling at supersonic and hypersonic speeds and are encountered, {\em e.g.,} at engine inlets, fin junctions and control surfaces on which complex supersonic flows with several shocks, shear layers and recirculation zones are created. The interaction of these flow structures are known to create so-called Edney-IV shock interactions~\citep{Edney} which are well-known to lead to unsteadiness in the flow  \citep{DollingreviewSWBLI,ClemensReviewSWBLI,GAITONDE201580}. Such flow characteristics can also trigger laminar to turbulent transition when the incoming boundary layer is laminar, as initially described in the work of \cite{chapman1958investigation}, the series of early experimental investigations of \cite{ginoux1960existence,ginoux1966laminar,ginoux1969some}, and the subsequent comprehensive experiments of \cite{SimeonidesPhDThesis}. Consecutively, transition location needs to assessed accurately as transition to turbulence enhances the heat transfer rate and introduces unsteadiness, both of which crucial for the considerations of range and control of the vehicle.

Hydrodynamic linear stability theory (LST) based on the linearized Navier-Stokes equations has been widely used to predict transition in supersonic flows \citep{Mack1969,Mack1984}. In the context of global LST \citep{Theofilis2003,theofilisARFM}, analysis of steady laminar two-dimensional supersonic flows has been used to predict linear instability characteristics in a wide range of spatially inhomogeneous high-speed flows including, for example, shock wave / flat plate laminar boundary layer interactions \citep{BoinEtAlTCFD,robinet_2007}, the wake of an isolated roughness element embedded in a laminar boundary layer \citep{DeTullioEtAlJFM2013}, as well as modal \citep{paredes_gosse_theofilis_kimmel_2016} and nonmodal \citep{quintanilha2022transient} instability of hypersonic flow over an elliptic cone at zero angle of attack.

Early experimental studies of compression ramps focused on overall properties of the flow, such as heat transfer rates and pressure distributions. \cite{Holden1966} and \cite{Holden1970THEORETICALAE} investigated the dependence of heat transfer rates on the separation and reattachment lengths and found that they strongly depend on the reattachment angle. \cite{Bloy_Georgeff_1974} performed experiments at Mach 12 on a compression ramp with a sharp leading edge and observed good agreement with theory \citep{KlinebergPhDThesis, KlinebergAIAAJ} for the pressure and heat transfer distributions. The appearance of the separation region as the ramp angle increases in a supersonic free stream was considered by \cite{Settles1979AIAAJ}, who also showed agreement with their Navier-Stokes simulations when no separation bubble was formed at the ramp corner. \cite{Andreopoulos_Muck_1987} also worked on several ramp angle configurations at Mach 3, however with a turbulent incoming boundary layer. Their work provided insights into the shock oscillations and concluded that the state of the incoming boundary layer, rather than any downstream effect, is most likely triggering this oscillation. Hypersonic flows over compression corners were investigated extensively {\color{black} in a series of experiments by} \cite{SimeonidesPhDThesis}, \cite{SimeonidesetalAIAAJ1994} and \cite{Simeonides_Haase_1995}. These authors considered Mach 14 free stream flow, a 15$^\circ$ ramp angle and several flat plate and ramp length configurations. Most interestingly they observed that, even though the incoming boundary layer was linearly stable and extremely hard to trip to turbulence, transitional flows were observed near the reattachment region, which these authors attributed to the adverse pressure gradients created by the separation region. Accurate experimental results that could be used to compare with the numerical studies were also the focus of \cite{Settles1994AIAAJ}, who identified experiments which could be used to compare with numerical work. In the experimental study of compression ramps of \cite{mallinson_gai_mudford_1997} the focus was on real gas effects at several high-enthalpy hypersonic conditions. At the range of examined enthalpies, $3-19 MJ kg^{-1}$, it was found that real gas effects were negligible. On the other hand, the experimental studies of \cite{Hozumi2001AIAA} centered their attention on the heat transfer rates at high compression ramp angles in Mach 10 flow as a function of the leading-edge bluntness. Rather than using different ramp angles, the angle of attack of the compression ramp was changed to create interaction regions of different strengths and it was observed that at high angle of attack different Edney-type interactions strongly affect the heat transfer rates. These authors also conducted numerical studies to compare with their experiments and showed that three-dimensional simulations gave better agreement with experiments than the corresponding two-dimensional work, which suggested an effect of three-dimensionality of the separation region. \cite{Ringuette2009AIAAJ} is another example of experimental investigation of supersonic flow over a nominally two dimensional compression ramp geometry at Mach 3 with a turbulent incoming boundary layer. These authors compared low- and high Reynolds number cases and showed that, in the high-Reynolds number limit, the separation bubble is larger and fluctuation components of the pressure have smaller root mean square values. 
\cite{GANAPATHISUBRAMANI_CLEMENS_DOLLING_2009} characterized experimentally the low-frequency unsteadiness of Mach 2 flow, again with a turbulent incoming boundary layer, and stated that low-frequency oscillations in the separation bubble are influenced by both local and global effects of the incoming boundary layer region.  \cite{Roghelia2017} experimented with several compression corner angles to gain insight into the so-called {\em G\"ortler vortices}, i.e. centrifugal instability which gives rise to streamwise-aligned structures. \textcolor{black}{They concluded  that these vortices increase the streamwise heat transfer rate. They also stated that even though the spanwise heat transfer rate did not strongly vary, the increase the spanwise variation increased as the ramp angle increased.}
The related axisymmetric "flare" / compression ramp configuration has also received intense attention, but will not be discussed here; readers with an interest in instability and transition in this class of flows are referred, for example, to the recent works of \cite{benitez_borg_scholten_paredes_mcdaniel_jewell_2023} and \cite{DavamiEtAlAIAA2024-0499}, both at Mach 6. 

From a theoretical point of view, use of triple-deck theory \citep{neiland1969theory,stewartson1970laminar,Messiter} has been shown to provide significant insights into integral quantities, unsteadiness and instability of compression corner flows. Early triple deck calculations were used to estimate the skin friction coefficient along the compression ramp and separation lengths at high Reynolds numbers \citep{rizzetta1978triple}. 
Along with predicting the mean flow characteristics, triple-deck formulations were also used to relate unsteadiness with linear instability \citep{cassel_ruban_walker_1995,fletcher_ruban_walker_2004} and show that this phenomenon occurs on account of an increase in the ramp angle, below which steady flow is obtained. One aspect that was missing from the early triple-deck studies is inclusion of the effects of the leading edge shock, which was first considered by \cite{CowleyHall1990} and \cite{SeddouguiBassom1997} in analyses of supersonic flows past a wedge and a sharp cone, respectively. The latter studies showed that inclusion of the attached leading edge shock modifies the decks of the flow and results in amplified three dimensional/non-axisymmetric viscous modes, especially when the shock is in close proximity to the boundary layer, i.e. the distance of the shock layer from the wall is comparable to the boundary layer thickness. One of the most interesting results of triple-deck theory is the correlation of unsteadiness in a compression corner with an appropriately scaled ramp angle \citep{NeilandBogolepovDudinLipatov}, a result that was verified  numerically using two- and three dimensional direct numerical simulation (DNS) under a variety of flow conditions \citep{egorov2011three}.  \cite{gai_khraibut_2019} have used the scaling laws of triple-deck theory to explore the dynamics of larger separation bubbles and the appearance of secondary recirculation regions within the laminar separation bubble using numerical solutions of the Navier-Stokes equations. They also showed that, at high enough Reynolds numbers at which triple-deck theory is expected to be valid, a secondary circulation in the separation bubble is expected for scaled angles higher than about four. On the other hand, the work of \cite{GrishamEtAl} has cast some doubt on the degree of agreement expected between triple-deck theory and numerical solution of the full compressible Navier-Stokes equations of motion; this open question will be addressed elsewhere.

Linear global stability analysis methods were also used to predict transition on supersonic compression ramp flows. In fact, most of the recent numerical and experimental studies on compression corner flows show that laminar flows become three dimensional and unsteady in the separation region, which then triggers transition to turbulence downstream of the compression corner. Reattachment streaks were shown by  \cite{dwivedi_etal_2019} to arise past the compression corner at a free stream Mach number of 8.0 and Reynolds number based on the flat plate length of $2.0-3.7\times10^{5}$ on account of the linear global instability associated with the stationary self-excited mode of laminar separation. This mode, henceforth termed the {\em C-shaped mode} on account of the structure of the spanwise perturbation velocity component, is topologically identical with that discovered by \cite{BoinEtAlTCFD} and \cite{robinet_2007} in the related supersonic shock impingement problem and, in turn, with their incompressible counterpart discovered by \cite{Theofilis2000} {\color{black} in a canonical steady, nominally two-dimensional laminar separation bubble}. In a series of recent publications, \cite{Cao_Hao_Klioutchnikov_Olivier_Wen_2021} studied unsteady effects of the hypersonic flow over a compression ramp using DNS and global LST and compared their result to their experiments at free-stream Mach number 7.7, Reynolds number $4.2\times10^{5}$ based on a (relatively long) flat plate length and a shallow $15^\circ$ ramp angle. It was shown that streamwise heat flux streaks appear downstream of the reattachment and the flow exhibits low frequency unsteadiness due to global instability. Their work continued on the same configuration and flow parameters by performing detailed three-dimensional global LST \citep{hao_occurancer_2021}, which showed that secondary recirculation regions appear as the ramp angle is increased. In follow-up work \cite{Cao_Hao_Klioutchnikov_Olivier_Heufer_Wen_2021} investigated the effect of leading edge bluntness on stability and showed that, as the leading edge bluntness increased up to a critical value, the size of the laminar separation bubble also increases, however after this critical value is exceeded, the bubble starts to shrink. Global stability analysis of the steady laminar two-dimensional fields showed that the amplification rate of the most unstable modes, manifested in the form reattachment streaks, followed the same trend, first increasing as the ramp angle increased and then decreasing past the critical ramp angle.   
\cite{dwivedi_etal_2022} performed resolvent analysis on flow over a double wedge at Mach 5 and free-stream unit Reynolds number of $13.6\times10^{6} m^{-1}$
and showed that external disturbances entering the flowfield upstream of the separation region result in oblique waves which give rise to three-dimensionality and laminar-turbulent transition.
At conditions close to those addressed in their earlier analyses and experiments, \cite{cao_etal_2022} showed that transition to turbulence downstream of the reattachment location was tripped by the reattachment streaks and that low-frequency unsteadiness was also present downstream of the reattachment region. The work of \cite{Cao_Hao_Guo_Wen_Klioutchnikov_2023} addressed the effect of different degrees of rounding the compression corner on instability of the laminar separation bubble, employing modal and non-modal (resolvent) global stability analysis and DNS. These authors showed that increasing the radius of curvature of the compression corner leads to damping of the instability and ultimately leads to elimination of separation. At certain spanwise periodicities amplification of streamwise streaks at low frequencies was shown, while the theoretical results were found to agree with experiments. 
In their most recent work, \cite{Hao_Cao_Guo_Wen_2023} employed global resolvent analysis for the conditions of \cite{Cao_Hao_Klioutchnikov_Olivier_Wen_2021} and addressed the effect of ramp angle on flow stability, increasing the ramp angle from zero, i.e. a flat plate. Their results showed that optimal response to the external forcing peaks at the leading edge and gives rise to streamwise streaks; in turn these get amplified in the separation bubble and become G\"ortler-like vortices.

To-date, at least two aspects of compression ramp linear flow instability have not been covered by the aforementioned activities. First, compression corners addressed in the literature are typically at relatively large distances from the leading edge of the flat plate preceding the ramp, such that a well-defined, laminar or turbulent, boundary layer is formed ahead of the separation zone; in this manner, the separation zone does not interact with the leading edge shock and shock wave / boundary layer interactions can be addressed in isolation from the leading-edge shock. Second, the recent demonstration of synchronization of the stationary three-dimensional global mode of the laminar separation bubble formed on the related configuration of flow over a double wedge with a previously unknown linear instability inside the (fully resolved in DSMC) separation shock layer \citep{sawant_etal_2022} raises the possibility of an analogous mechanism being present in compression ramp flows, when the plate length is short enough and the ramp angle is sufficiently large for the separation to interact with the leading-edge shock.

The present contribution is motivated by open questions arising in {\color{black} the latter} context. First, a relatively short length of the flat plate that precedes the compression corner is chosen, in conjunction with large ramp angles compared with those typically found in the literature. Second, the laminar steady state is computed by Direct Simulation Monte Carlo methods, at a Knudsen number that {\color{black} is low enough to be consistent with} subsequent use of linear stability analysis equations based on the continuum assumption. Section \ref{sec:3DFlow} introduces the large-scale numerical work performed using DSMC to compute two- and spanwise-periodic, three-dimensional flows over short compression corners at two ramp angles. The physical and numerical parameters of the problem {\color{black} are introduced} and the {\color{black} main} findings of the simulations {\color{black} are presented.}  Linear stability analysis is performed in Section \ref{sec:BiGlobalStability} to explain {\color{black} these} findings; besides the stationary three-dimensional global mode  known from analyses of compression corners attached to long flat plates, which is stable at all conditions analyzed, a new traveling global mode is discovered, associated with the interaction of the leading edge shock and the shear layer formed on the large separation bubble that extends to the lip of the flat plate. Section \ref{sec:Unsteady42Ramp} presents full three-dimensional (unsteady, nonlinear) DSMC simulations in which the linear stability analysis results are fully verified and the evolution of the flow on the ramp is followed during the early nonlinear stages of formation of lambda-vortices that ensue linear modal growth. Concluding remarks are offered in Section \ref{sec:conclusions}.

\section{DSMC computations of short three-dimensional compression ramps}\label{sec:3DFlow}

In the present work, supersonic two- and (spanwise-periodic) three-dimensional flows over compression ramps at two angles of 30$^{\circ}$ and 42$^{\circ}$ are computed using Direct Simulation Monte Carlo methods~\citep{Bird} and the  Scalable Unstructured Gas-dynamic Adaptive mesh Refinement (SUGAR) solver~ \citep{Sawant2018}.  The free-stream parameters are the same for both simulations and are given in Table~\ref{tab:FreeStreamDSMC} for a  working fluid of molecular nitrogen. The geometry consists of a  flat plate and ramp which begin at $x=0.01$~m and $0.19$~m, respectively, with a domain length  of $0.30$~m for both 2D and 3D configurations and with domain heights of $0.25$ and $0.10$~m, for 2D and 3D configurations, respectively. \textcolor{black}{The length of the plate, $L=0.18$~m, is used as length scale in the analyses.}
 The flat plate and  ramp walls are assumed to be isothermal  with a  temperature of $273~K$. In terms of DSMC numerical parameters for the 2D simulations, the number of simulation particles in the flow domain is about $ 45\times10^{6}$ with at least 50 particles in each sampling cell.   Spanwise periodic 3D DSMC simulations (designated as 3D-SP) were performed using spanwise lengths suggested by linear three-dimensional global (BiGlobal) stability analysis.  As will be shown in Sec.~\ref{sec:BiGlobalStability}, a spanwise length of $\lambda = 0.56 L$  gives a  spanwise wavenumber of the least stable mode, $\beta = 11$.  
The sampling cell size and time step were kept the same as in the 2D simulations, as given in Table~\ref{tab:FreeStreamDSMC}. A total of  $30 \times 10^{6}$ sampling cells and $15.1 \times 10^{9}$ simulation particles were used in the 30$^{\circ}$ three-dimensional simulations, while $57.6 \times 10^{6}$ sampling cells and $30.2 \times 10^{9}$ simulation particles were used in the corresponding 42$^{\circ}$ simulations.
 Moreover, it was verified that there were at least 20 particles per collision cell in order to obtain sufficient statistics and accurately capture the time behavior of the relevant {\color{black} macroscopic flow quantities}.  
To model gas particle collisions, the majorant frequency scheme~\citep{majorant} was implemented in SUGAR. The viscosity-temperature dependence of the gas was accurately modeled using the Variable Hard Sphere (VHS)~\citep{Bird} model for molecular nitrogen at a reference temperature of $273~K$, \textcolor{black}{with molecular diameter and viscosity coefficient of $4.17\times10^{-10}$~m and $0.255$, respectively.} Since  one of the methodologies  used to study the nature of the compression ramp flow is BiGlobal linear stability analysis, two-dimensional steady base flows were obtained by sampling for 50,000 time steps (equivalent to 2.5~ms) after the (unsteady) simulations converged to a steady state. 

Triple-deck theory~\citep{neiland1969theory,stewartson1970laminar,Messiter} suggests a non-dimensional scaled angle parameter, $\alpha^{*}$, to characterize the compression ramp flows, and defined as the ratio of the wall normal and streamwise length scales,
\begin{equation}
    \alpha^{*} = \alpha \frac{Re^\frac{1}{4}}{0.332^\frac{1}{2}C^\frac{1}{4}(M_\infty^2 - 1)^\frac{1}{4}}, \hspace{20pt} C = \frac{\mu_w T_\infty}{\mu_\infty T_w}  \label{eq:alpha}
\end{equation}
where $\alpha$ is the angle of the ramp in radians, and subscripts $w$ and $\infty$ denote values at the wall and freestream, respectively. This non-dimensional parameter will be used to characterize the stability of the separation bubbles in the two dimensional analyses.

\begin{table}
\caption{Free stream conditions and DSMC computational parameters}
\label{tab:FreeStreamDSMC}
\begin{center}
\scalebox{0.8}{
\begin{tabular}{ccccc}
\hline
\hline
Parameter&Value \\
\hline
Velocity [m/s], $U_\infty$& 432\\
Temperature [K], $T_\infty$& 50\\
Wall Temperature [K], $T_w$& 273\\
Number Density [m$^{-3}$], $n_\infty$& $ 14.32\times10^{21}$  \\
Mach Number, $M_\infty$& 3.0 \\
Reynolds Number, Re$_L$& $11.2\times10^{3}$  \\
Flat Plate Length [m], L& 0.18\\
Knudsen Number, Kn$_L$ & $3.0\times10^{-4}$\\
Sampling Cell Size [m]& $ 2.0\times10^{-4}$ \\
Time Step [s]& $ 5.0\times10^{-8}$ \\
Flow Time [s], ${\tau _{flow}}=L/{U_\infty }$ & $4.17\times10^{-4}$\\
\hline
\hline
\end{tabular}}
\end{center}
\end{table}

\begin{figure}
\center
{\includegraphics[trim=200 40 200 40,clip,width=0.6\linewidth]{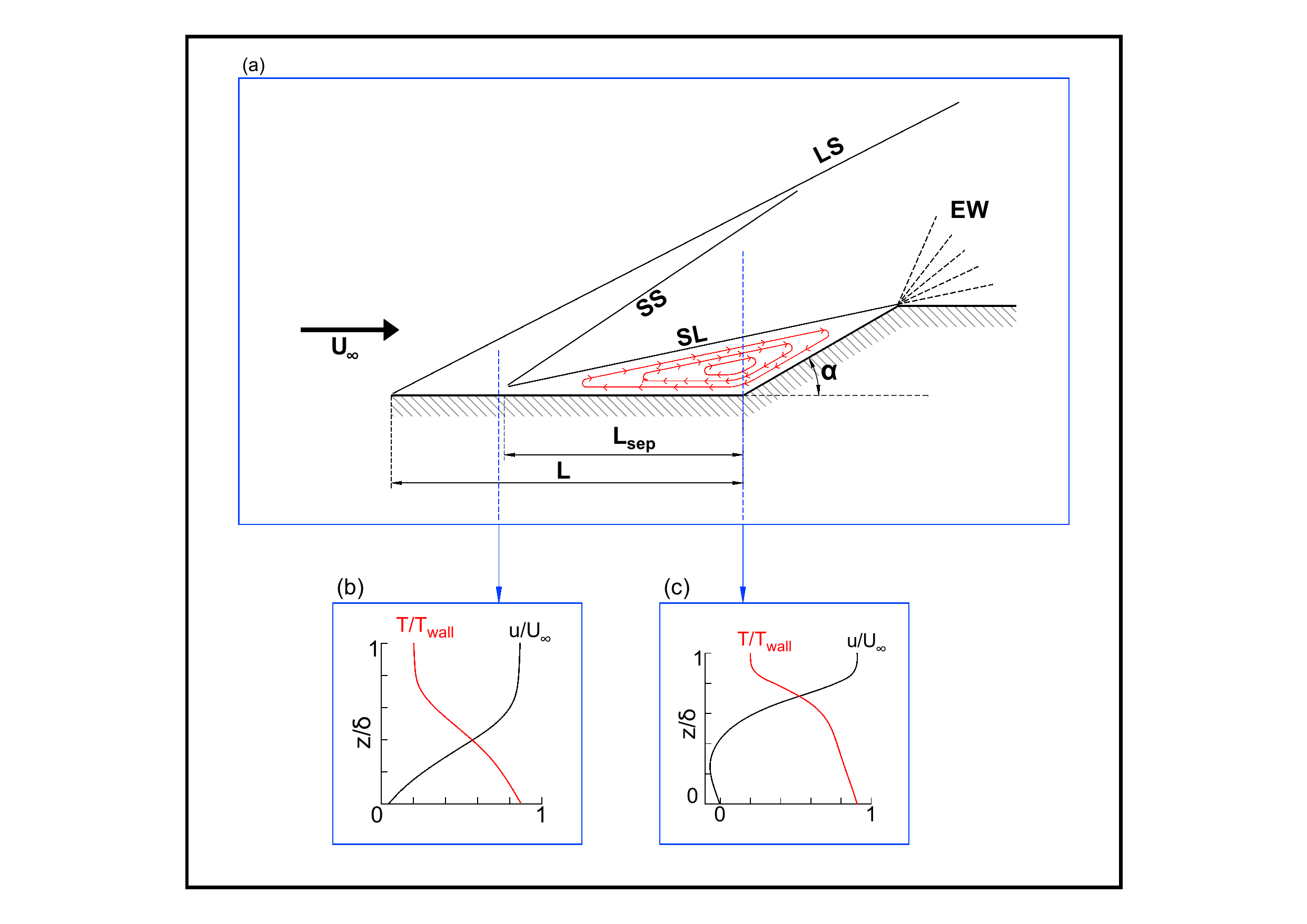}}
\caption{Flow structures for a canonical flow over a compression-expansion corner. (a) overall flow structures, (b) and (c) boundary layer profiles before and after the flow separation. LS: leading edge shock, SS: separation shock, SL: separated shear layer, EW: expansion waves. L and L$_{sep}$ represents flat plate length and separation length and $\alpha$ is the physical ramp angle.}
\label{fig:OverallFlowSketch}
\end{figure}

The flow over a compression ramp is a canonical flow that creates several characteristic structures, as shown in Fig.~\ref{fig:OverallFlowSketch}(a). First the leading edge of the flat plate creates an attached Leading-edge Shock (LS). Downstream, at sufficiently large ramp angles, there may exist a shock induced separation bubble, with an associated Separation Shock (SS) and separated Shear Layer (SL) and after the flow reattaches, Expansion Waves (EW) are present at the expansion corner. The lengths $L$, $L_{sep}$ and the ramp angle $\alpha$ that are used in the scaled angle and Reynolds number calculations also are shown in the Fig.~\ref{fig:OverallFlowSketch}(a). The local boundary layer thickness is indicated by $\delta $ in Figs.~\ref{fig:OverallFlowSketch}(b) and \ref{fig:OverallFlowSketch}(c). The  streamwise, spanwise and normal spatial directions are indicated by $x$, $y$, and $z$, such that the periodic side-boundaries in our three-dimensional simulations are along the $y-$direction. For the flow conditions considered in this work, our two-dimensional DSMC simulations have revealed that the boundary layer profile just upstream of separation still exhibits a developing profile, with slip velocity and temperature jump present, \textcolor{black}{as shown in the detailed view of Fig.~\ref{fig:OverallFlowSketch}(b). Within the separation bubble, the magnitude of the temperature jump has been reduced, the velocity profile shows negative values due to the flow reversal and at the wall the velocity is zero, i.e. no slip has been attained, also shown in the detailed view of Fig.~\ref{fig:OverallFlowSketch}(c).}

\begin{figure}
\center
\subfigure[]{\label{fig:302DTimeEvo}\includegraphics[trim=80 20 80 20,clip,width=0.40\linewidth]{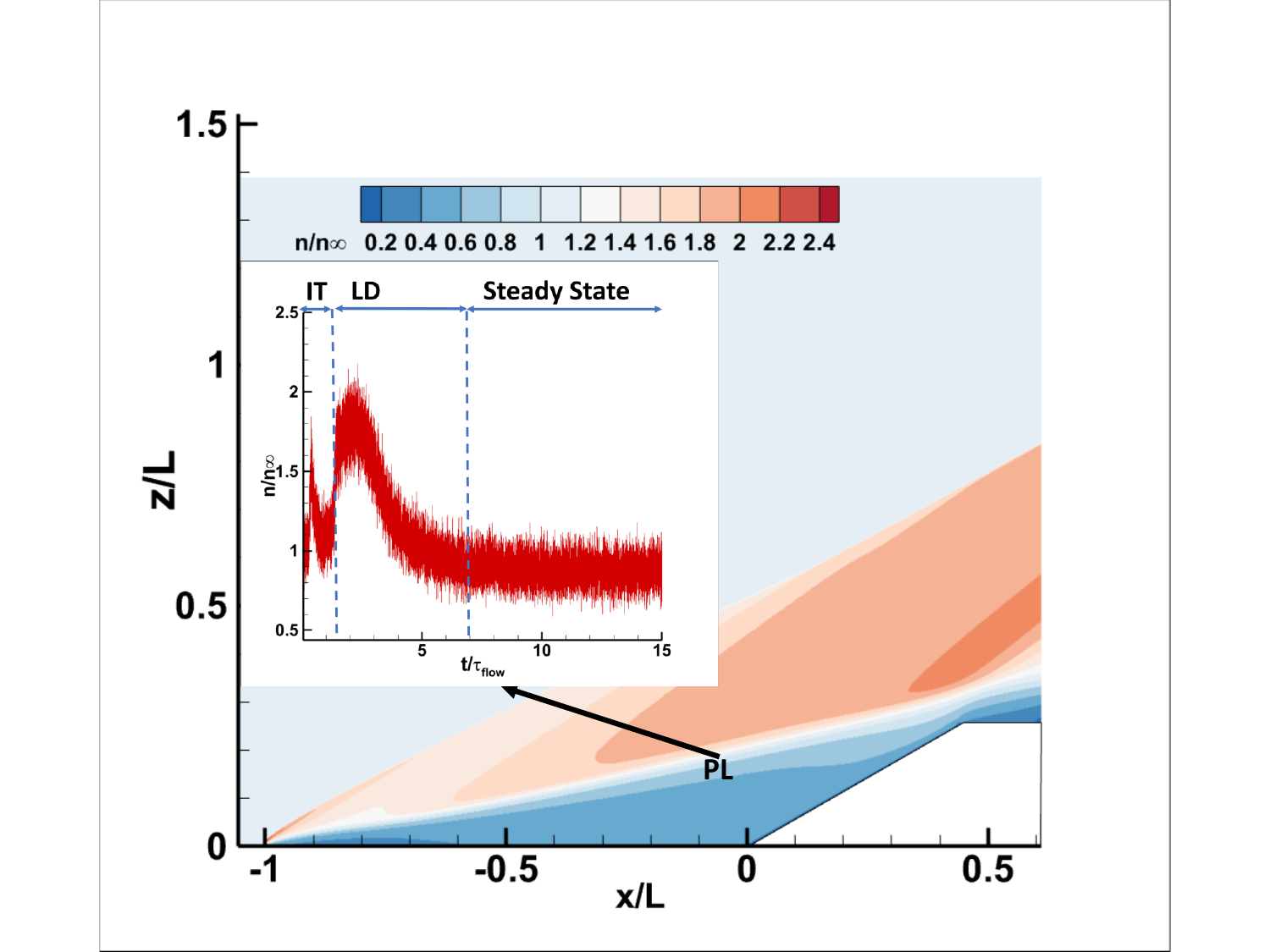}}
\subfigure[]{\label{fig:422DTimeEvo}\includegraphics[trim=80 20 80 20,clip,width=0.40\linewidth]{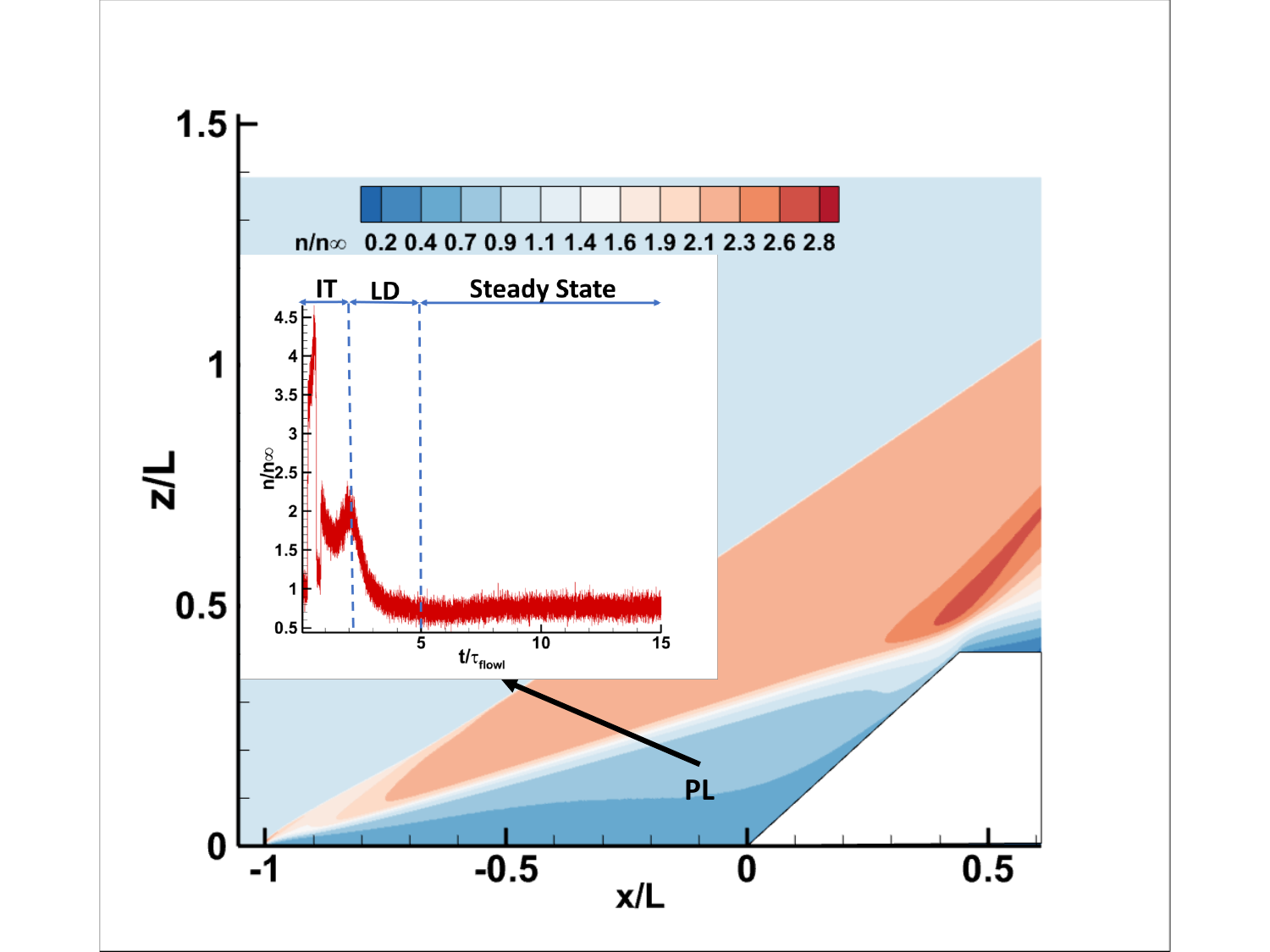}}
\caption{Normalized number density contours and time evolution of number density for (a) 30$^{\circ}$  and (b) 42$^{\circ}$ ramp angles. IT: Initial Transient, LD: Linear Decay and PL indicates the numerical probe location in the figures. $t_{flow}$ is defined in Table~\ref{tab:FreeStreamDSMC}. }
\label{fig:NdenContourTimeEvo2D}
\end{figure}

The flow number density contours of the time-averaged steady state results for the two-dimensional simulations as well as the time evolution inside the separation bubble for both the 30$^{\circ}$ and 42$^{\circ}$ compression ramps 
can be seen in Figure~\ref{fig:NdenContourTimeEvo2D}. Of these, the 30$^{\circ}$ angle configuration is closer to the compression ramp geometries  previously reported in the literature, 
where the separation length is comparable to that of the flat plate.  By contrast,  at 42$^{\circ}$ the flow is almost immediately separated downstream of the tip.  The time evolution of number density at both angles shows similar behavior with an initial transient (IT)
and a linear decay (LD) period, indicated on the figure. During the initial transient, shocks pass through as the flow establishes, while during linear decay exponential convergence of all flow quantities to their respective steady state values is observed. These results are qualitatively analogous to those obtained in DSMC simulations on related configurations at low Reynolds number values \citep{tumukluPoF2,TumukluPRF2019}.

\begin{table}
\caption{Comparison of streamwise extent of the separation bubble, scaled angle $\alpha^*$ and maximum bubble recirculation in two-dimensional DSMC simulations at 30$^{\circ}$  and 42$^{\circ}$}
\label{tab:SepBubbleComp}
\begin{center}
\scalebox{1.0}{
\begin{tabular}{ccccc}
\hline
\hline
$\alpha$ &$L_{sep}/L$& $\alpha^{*}$, using $L$ & $\alpha^{*}$, using $L-L_{sep}$ &  $\frac{|\min(U_x)|}{U_\infty}$ \\
\hline
30$^{\circ}$ &0.62& 6.1&4.1&10.5\%\\
42$^{\circ}$ &0.86& 8.6&5.7&16.3\%\\
\hline\hline
\end{tabular}}
\end{center}
\end{table}

\begin{figure}
\center
{\includegraphics[trim=70 30 60 30,clip,width=0.52\linewidth]{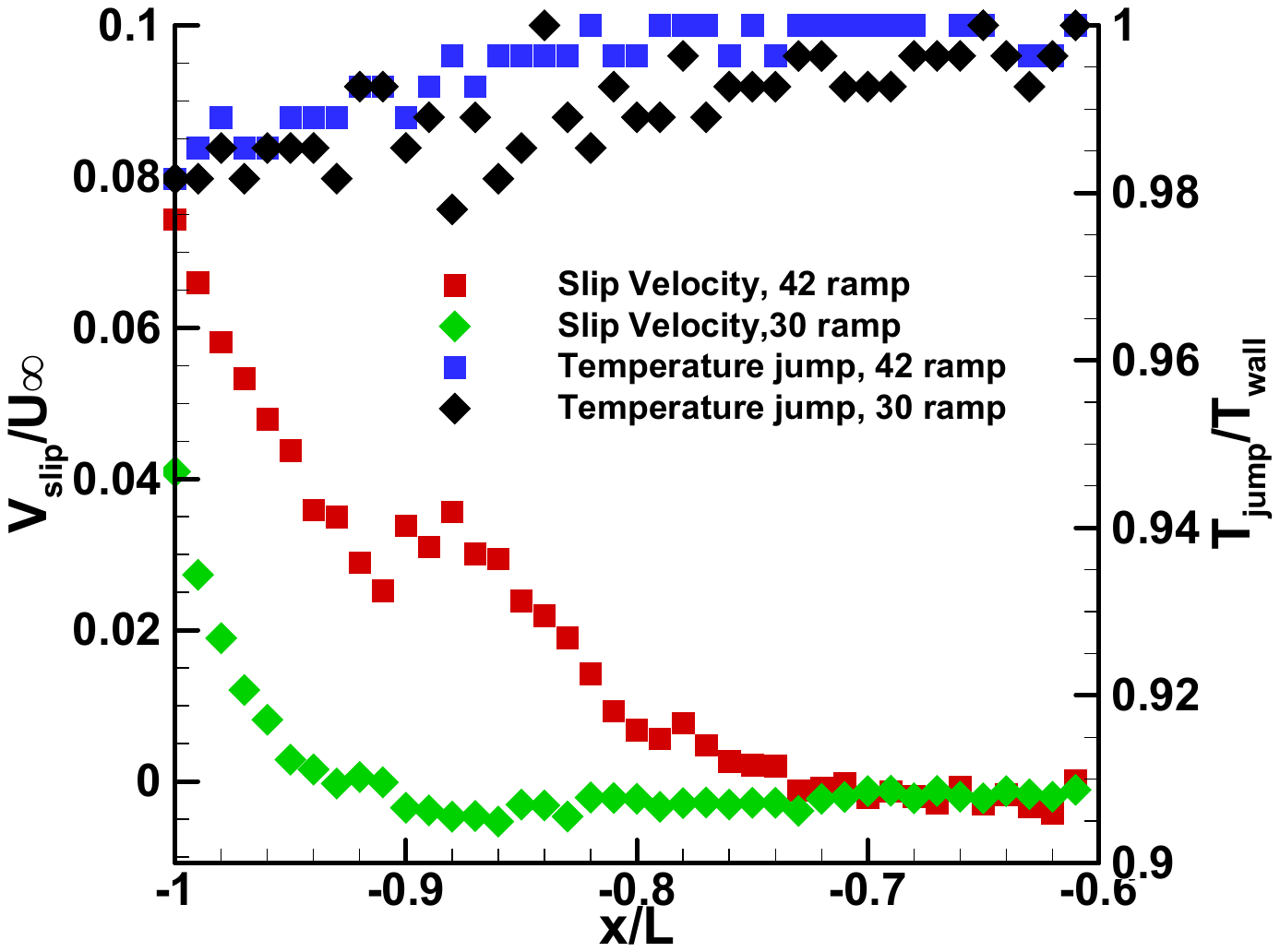}}
\caption{Velocity slip and temperature jump at the flat plate wall for 30$^{\circ}$  and 42$^{\circ}$  ramp angles. Separation occurs at $x/L =$ -0.62 and -0.86 at 30$^{\circ}$ and 42$^{\circ}$, respectively. }
\label{fig:TempjumpVelSlip}
\end{figure}

Although the triple-deck scaled angles  $\alpha^{*}$, computed with either $L$ or $L-L_{sep}$ and shown in Table~\ref{tab:SepBubbleComp}, are relatively high, 
no secondary vortices inside the bubble, indicative of unsteadiness and potentially global instability, appear at either angle. This may be explained by the effect of slip velocity and  temperature jump, rarefaction features that are also present at these free-stream continuum flows. Fig.~\ref{fig:TempjumpVelSlip} shows that the finite wall-velocities and temperatures are different at the two angles, until flow separates.  Slip velocity and temperature jump are computed as surface properties, sampling the gas particle speeds before and after they hit the surface over time. Downstream of separation, as the flow stagnates both macroscopic flow quantities come into equilibrium with the wall conditions, i.e. no-slip and isothermal.  At this moderate Reynolds number both of velocity slip and temperature jump are present in the vicinity of the leading edge and, as shown by \cite{INGER200742,Inger2008}, relax along the plate, allowing the flow to maintain a single recirculation structure.  

In selecting the compression ramp geometry values based on the scaling relationships of equation (\ref{eq:alpha}), one assumes a more developed boundary layer, that is not really present {\color{black} in our simulations}, especially {\color{black} not} at 42$^{\circ}$ ramp angle.
 Instead, when the length for the part of the flow that is attached is used, as given in the third column of Table~\ref{tab:SepBubbleComp},  a single recirculation is better justified according to the range of scaled angle values reported by \cite{gai_khraibut_2019} and \cite{hao_occurancer_2021}, where both values fall below 6.0. 
  Another parameter which provides guidance as regards to linear stability of the separated flow is the recirculation strength,  defined by the ratio of the negative maximum streamwise velocity to the free stream velocity.  As shown in Table~\ref{tab:SepBubbleComp}, values of 10.5\% and 16.3\% at 30$^{\circ}$ and 42$^{\circ}$ were obtained,  respectively. Recirculation values higher than 10\% have been found to lead to self-excited stationary 3D perturbations of laminar separation bubbles at supersonic conditions~\citep{BoinEtAlTCFD}, however, as will be shown shortly in Section~\ref{sec:BiGlobalStability}, instability in the present configuration does not originate at the separation bubble.

An additional aspect is revealed by analysis of the two-dimensional DSMC simulation results, namely that the leading edge shock exhibits continuum breakdown according to the parameter 
 \begin{equation}
\label{eqn:BreakdownParameter}
P = \sqrt {\frac{{\pi \gamma }}{8}} \frac{{M\lambda }}{\rho }\left| {\frac{{d\rho }}{{d\overline x }}} \right|,
\end{equation}
defined by \cite{Bird}, where {\color{black} values} $P>0.02$ computed presently inside the shock layer indicate continuum breakdown. This shows that even though the free stream conditions {\color{black} satisfy the continuum assumption}, continuum breakdown will occur in high gradient regions, such as the leading edge shock and \textcolor{black}{the near downstream region.}

\begin{figure}
\center
\subfigure[]{\label{fig:Sch302D}\includegraphics[trim=60 10 70 250,clip,width=0.43\linewidth]{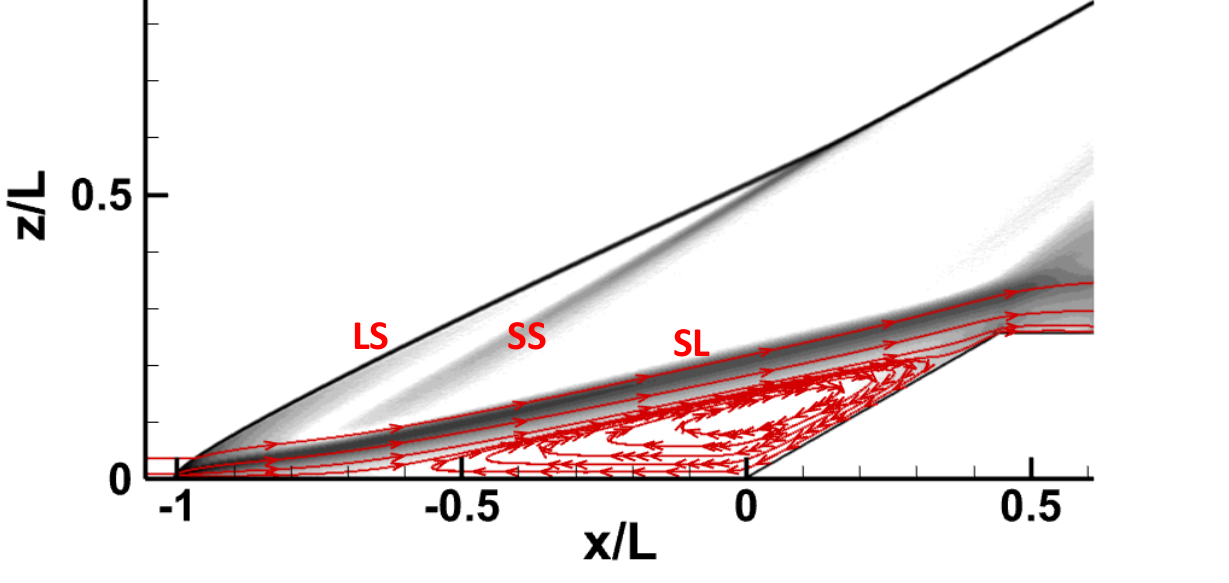}}
\subfigure[]{\label{fig:Sch303D}\includegraphics[trim=60 10 60  250,clip,width=0.43\linewidth]{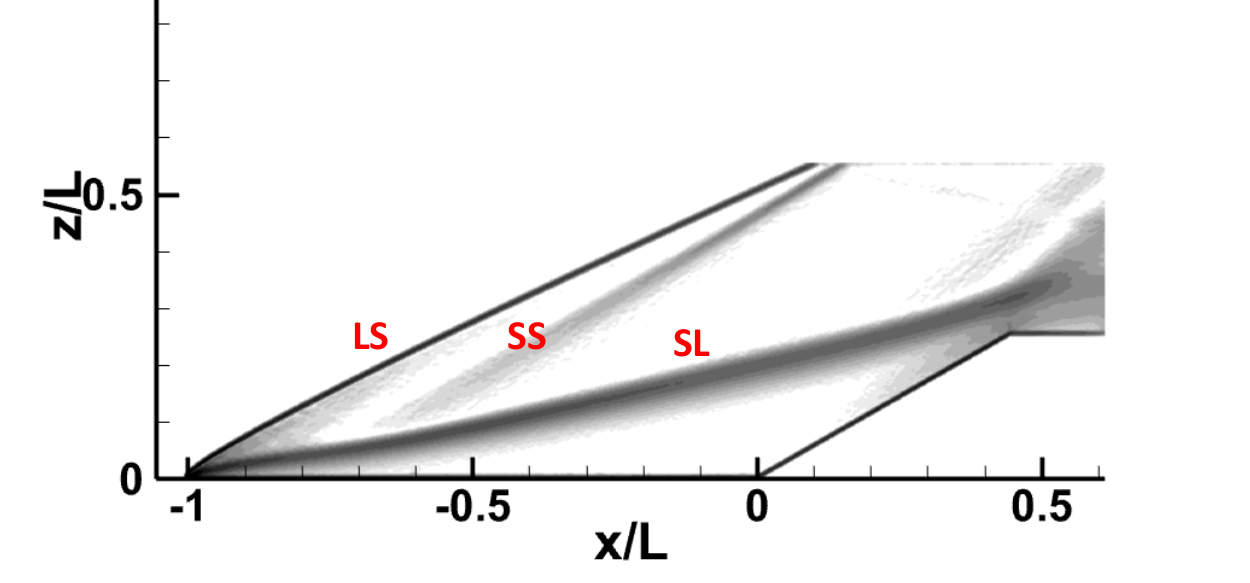}}
\subfigure[]{\label{fig:Sch422D}\includegraphics[trim=60 10 60  250,clip,width=0.43\linewidth]{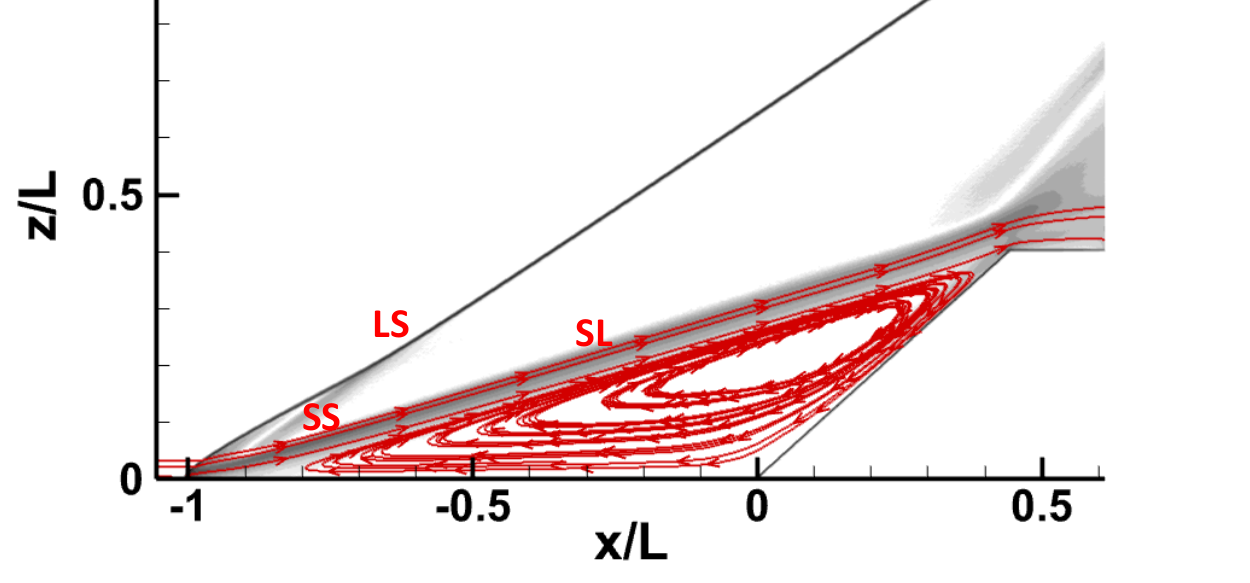}}
\subfigure[]{\label{fig:Sch423D}\includegraphics[trim=60 10 60  250,clip,width=0.43\linewidth]{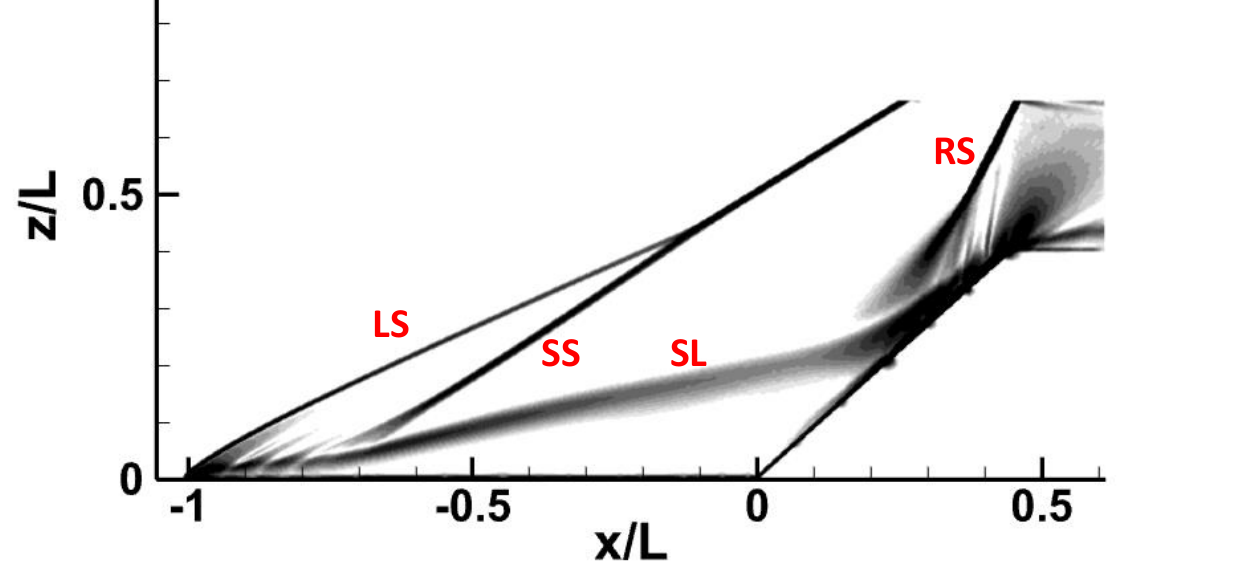}}
\caption{Schlieren images for the 2D (left column, (a),(c)) and 3D (right column, (b),(d)) simulations of the 30$^{\circ}$ (upper row, (a),(b)) and 42$^{\circ}$ (lower row, (c),(d)). For the 3D results middle plane cut along spanwise direction is shown. LS: leading edge shock, SS: separation shock, SL: separated shear layer, RS: reattachment shock. }
\label{fig:Sch2Dvs3D}
\end{figure}

Next, flow structures appearing in the two- and three-dimensional DSMC {\color{black} simulation results} are compared at  both ramp angles. Figure~\ref{fig:Sch2Dvs3D} shows the steady state numerical Schlieren images with streamlines, where it can be seen that the separation bubble covers a large portion of the flow region on the compression ramp at both ramp angles.  However, comparisons of the {\color{black}2- and 3D flows show different behavior at the two ramp angles}. At 30$^{\circ}$, both 2D and spanwise-periodic 3D simulations result in the same flowfield structures, with the leading edge shock (LS), separation shock (SS) and shear layer (SL) occurring at the same locations and flow separation starting at  $L_{sep}/L=0.62$, as seen in Figs.~\ref{fig:Sch302D} and ~\ref{fig:Sch303D}.  By contrast, {\color{black}comparisons at 42$^{\circ}$} in Figs.~\ref{fig:Sch422D} and ~\ref{fig:Sch423D} show qualitative differences in the flow features obtained in {\color{black} 2-} and {\color{black}3D} simulations. First, in 3D the separation bubble is smaller and the separation shock is stronger than in 2D. Second, since in {\color{black}3D flow reattaches earlier than in the 2D simulation}, a strong reattachment shock (RS) forms. Both of these indicate that the flow becomes essentially three-dimensional {\color{black}in the spanwise-periodic 3D simulations} and raise the question regarding the origin of this three dimensionalization. Before turning our attention to answering this question, it has been {\color{black}confirmed that, when the spanwise domain length is doubled in the 3D simulations, from $\lambda=0.56L$ to $1.12L$, the number of structures appearing in the flow is also doubled}. In fact, this doubling in the number of structures was only observed at 42$^{\circ}$ angle, while \textcolor{black}{no such variation was observed at 30$^{\circ}$, further confirming that the flow remains two-dimensional at the lower ramp angle.} In the remainder of the paper our attention is {\color{black} focused on} the ($\alpha=42^{\circ}$, $\lambda=1.12L$) case, in an effort to understand the origin of unsteadiness and three-dimensionalization of the flow in this nominally two dimensional geometry.

\section{Global linear stability analyses}\label{sec:BiGlobalStability}

The question of the origin of the  differences {\color{black} in the results of 2- and 3D simulations at the higher ramp angle (and the absence of such differences at the lowed angle) are} examined by application of linear modal global stability analysis to the {\color{black} respective} two-dimensional steady-state flows. Linear decomposition of fluid quantities into a basic, time-independent state, $\Bar{\mathbf{q}}$, and a small-amplitude {\color{black} unsteady} perturbation, $\Tilde{\mathbf{q}}$, is considered, according to
\begin{equation}
    \mathbf{q}(\mathbf{x},\tau) = \bar{\mathbf{q}}(\mathbf{x}) + \epsilon \tilde{\mathbf{q}}(\mathbf{x},\tau), \hspace{20pt}  with \hspace{10pt} \epsilon \ll 1.
\label{eqn:linearise}
\end{equation}
Here $\mathbf{x}$ is the normalized spatial coordinate vector ($x/L,y/
L,z/L$) and $\tau$ is the normalized time ($t/\tau_{flow}$). Assuming a homogeneous spatial direction $y$, along which the structure is periodic, the BiGlobal ansatz is used,
\begin{equation}
\tilde{\mathbf{q}}(\mathbf{x},\tau)= \hat{\mathbf{q}}(x,z)e^{i(\beta y - \omega \tau)}.
\label{eqn:purt}
\end{equation}
Here $\beta$ is the imposed spanwise wavenumber and $\omega$ is the complex eigenvalue sought in the solution of the temporal eigenvalue problem. The real and imaginary components denote the angular frequency and the  the growth or damping of linear perturbations, with negative values of the latter quantity denoting exponential decay of perturbations. The corresponding $\hat{\mathbf{q}}(x,z)$ is the vector of amplitude functions $(\hat{u},\hat{v},\hat{w},\hat{T},\hat{p})^T$ of the perturbations and the spanwise wavenumber and spanwise periodicity length are related by, 
\begin{equation}
\label{eqn:Spanlength}
\lambda ^*  = \frac{{2\pi }}{\beta } = \frac{{{\lambda}}}{L}
\end{equation}

\begin{figure}
\center
{{\includegraphics[trim=80 10 80 10,clip,width=0.68\linewidth]{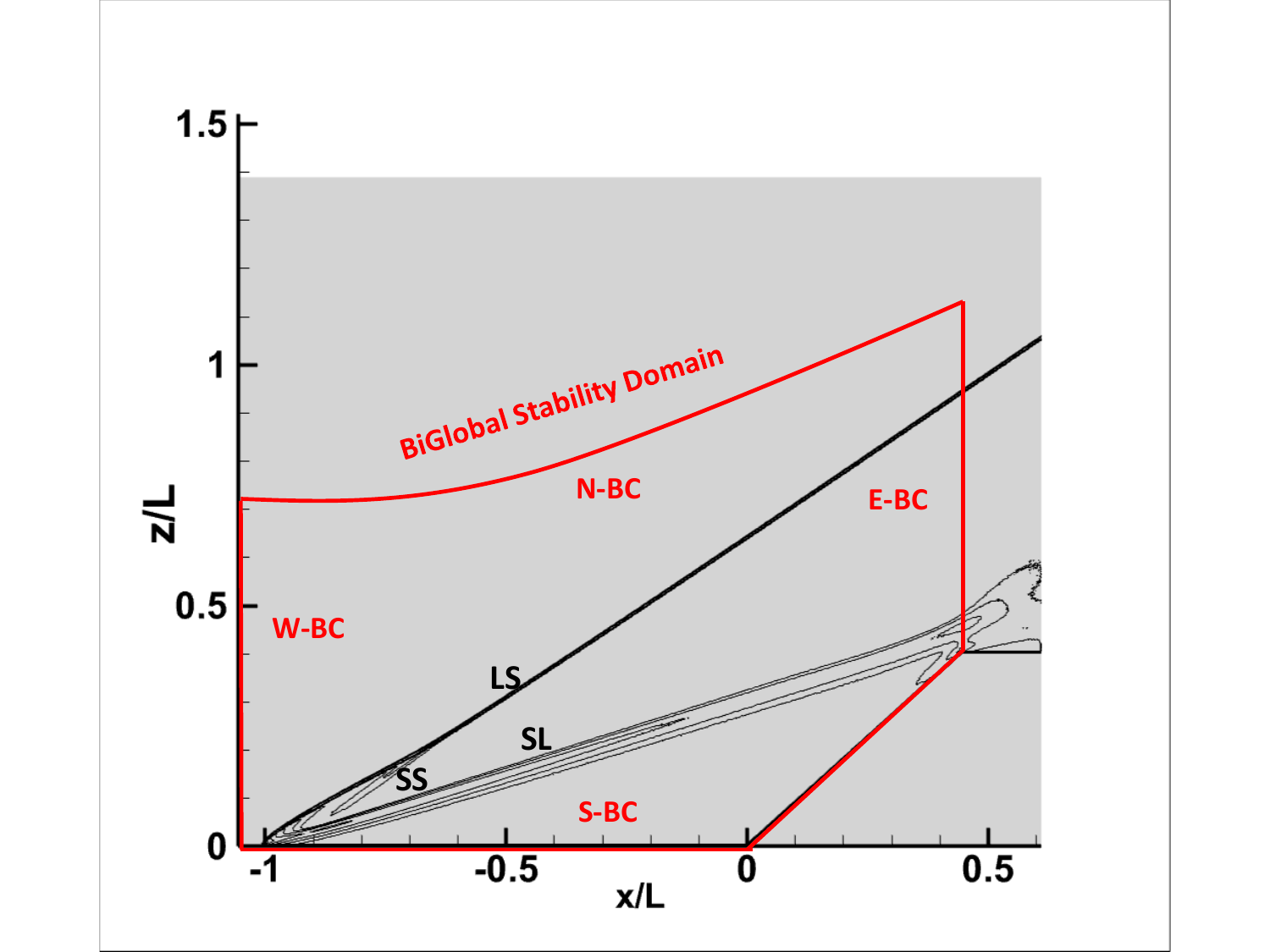}}}
\caption{Setup for the stability analyses superposed upon the numerical Schlieren of the base state and indicating the imposed boundary conditions (BC). W-BC: Dirichlet, S-BC: wall, E-BC: extrapolation, N-BC: extrapolation.}
\label{fig:LIGHTsetup}
\end{figure}

The in-house {\color{black} {\em LiGHT} (Linear Global Hypersonic Transition), generalized-coordinate, spectrally-accurate, modal and non-modal instability} analysis code \citep{theofilis2020massively,quintanilha2022transient} was used for this study.   The two-dimensional domain in which analyses are performed is shown in Fig.~\ref{fig:LIGHTsetup}, where it can be seen that the base flow LE shock is included and fully resolved.  
This is the main difference of the present work, compared to the earlier linear stability analyses of compression corners discussed in the Introduction, none of which included the LE shock, consistent with the fact that in the literature configurations the leading edge is far from the compression corner and the separation length is a fraction of the flat plate length. However, since most of the high gradient regions for the flows considered in this work concentrate towards the flat plate leading edge, as well as the dual fact that the shock itself creates unsteadiness ~\citep{sawant_POF,sawant_TCFD} and linear instability may extend in the shock layer region~\citep{sawant_etal_2022}, an analysis domain in which all the high gradient regions including the LE shock was chosen. 
Since the base flows were computed using DSMC, all gradients have been fully resolved and the DSMC thermal fluctuation was minimized by time averaging the flowfield over long integration times. Base flows were computed on a Cartesian grid and then interpolated onto a boundary fitted grid, as described in \cite{pagella2004}.  Subsequently, the physical domain was transformed into the numerical grid on which the eigenvalue problem was solved; full details of the numerical procedure can be found in \cite{CerulusPhD}.  {\color{black} At the wall Dirichlet boundary conditions have been used on the disturbance velocity components and disturbance temperature, while Neumann conditions were imposed on the disturbance pressure.} Dirichlet conditions have been applied on all amplitude functions at the inlet to ensure that no perturbations are permitted to enter the domain, while extrapolation is used both at the north and outlet domain boundaries.

\begin{figure}
\center
\subfigure[]{\label{fig:spectra30}\includegraphics[trim=80 10 80 10,clip,width=0.48\linewidth]{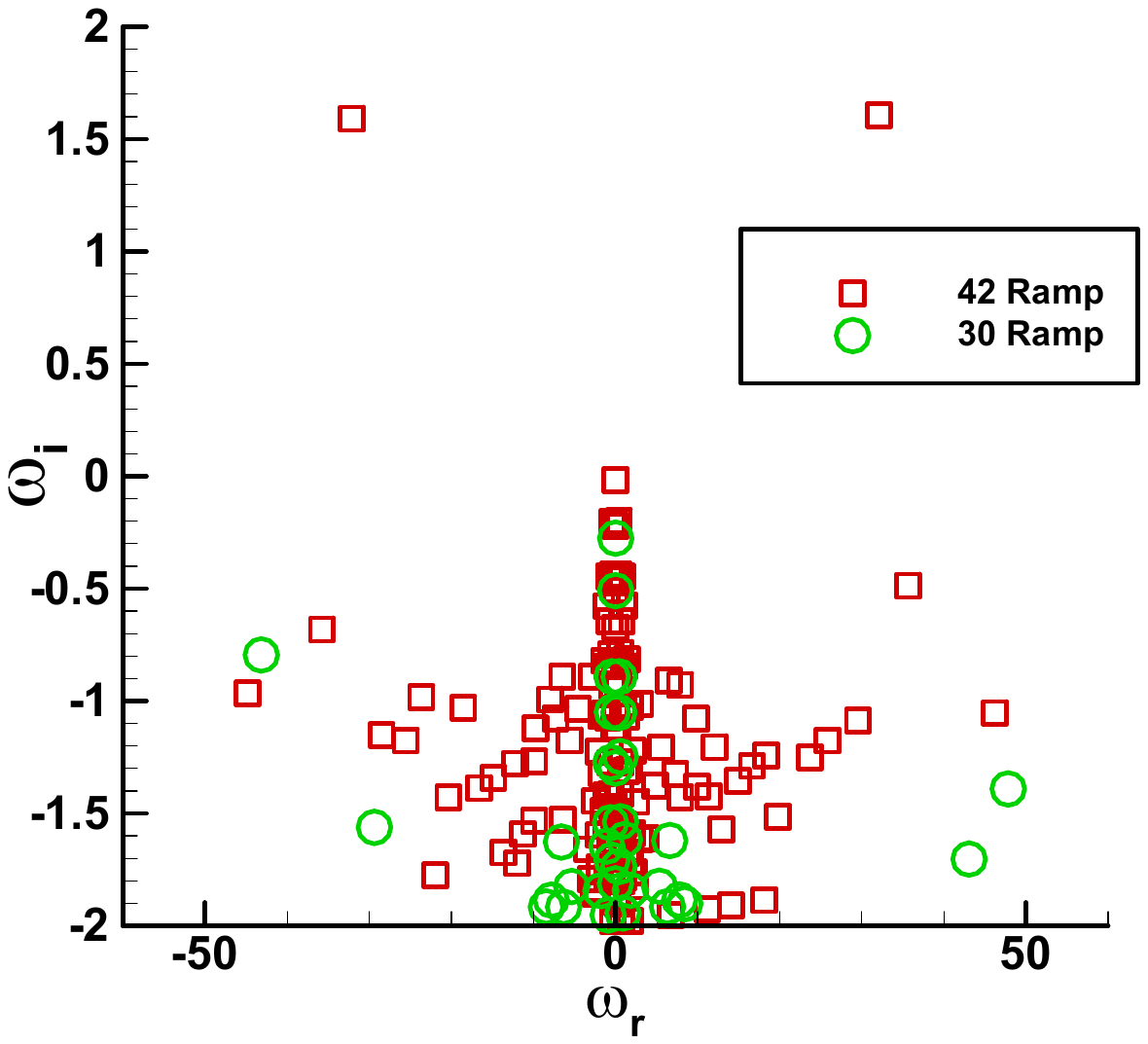}}
\subfigure[]{\label{fig:spectra42}\includegraphics[trim=80 10 80 10,clip,width=0.48\linewidth]{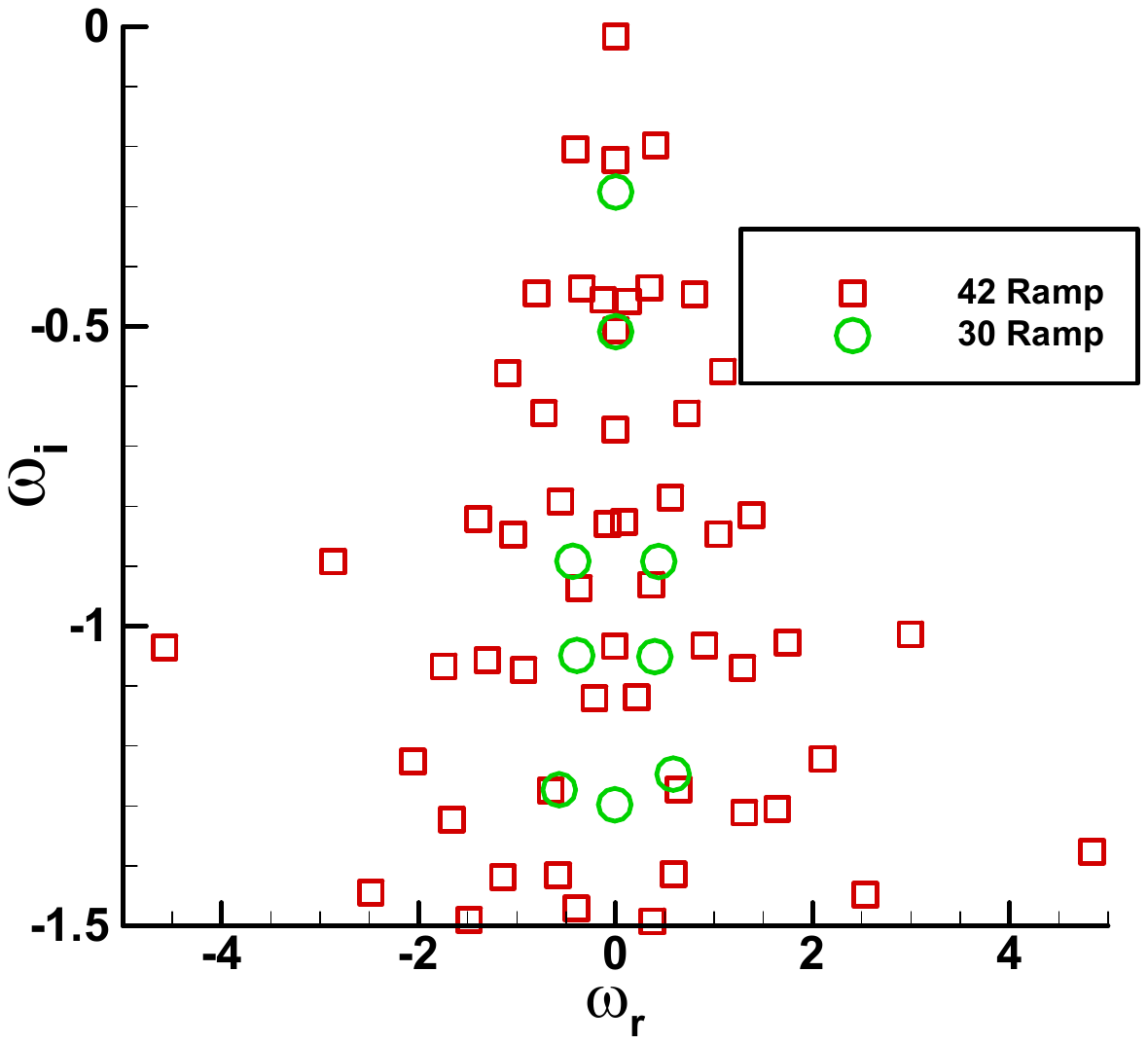}}
\caption{(a) Eigenvalue spectrum at $\beta=11$ for 30$^{\circ}$ (green circles)  and 42$^{\circ}$ (red squares) angle cases. In (b) a detailed view of the spectra is shown for stable modes.}
\label{fig:Evals3042}
\end{figure}

\begin{figure}
\center
\subfigure[]{\label{fig:Mode130}\includegraphics[trim=80 10 80 250,clip,width=0.48\linewidth]{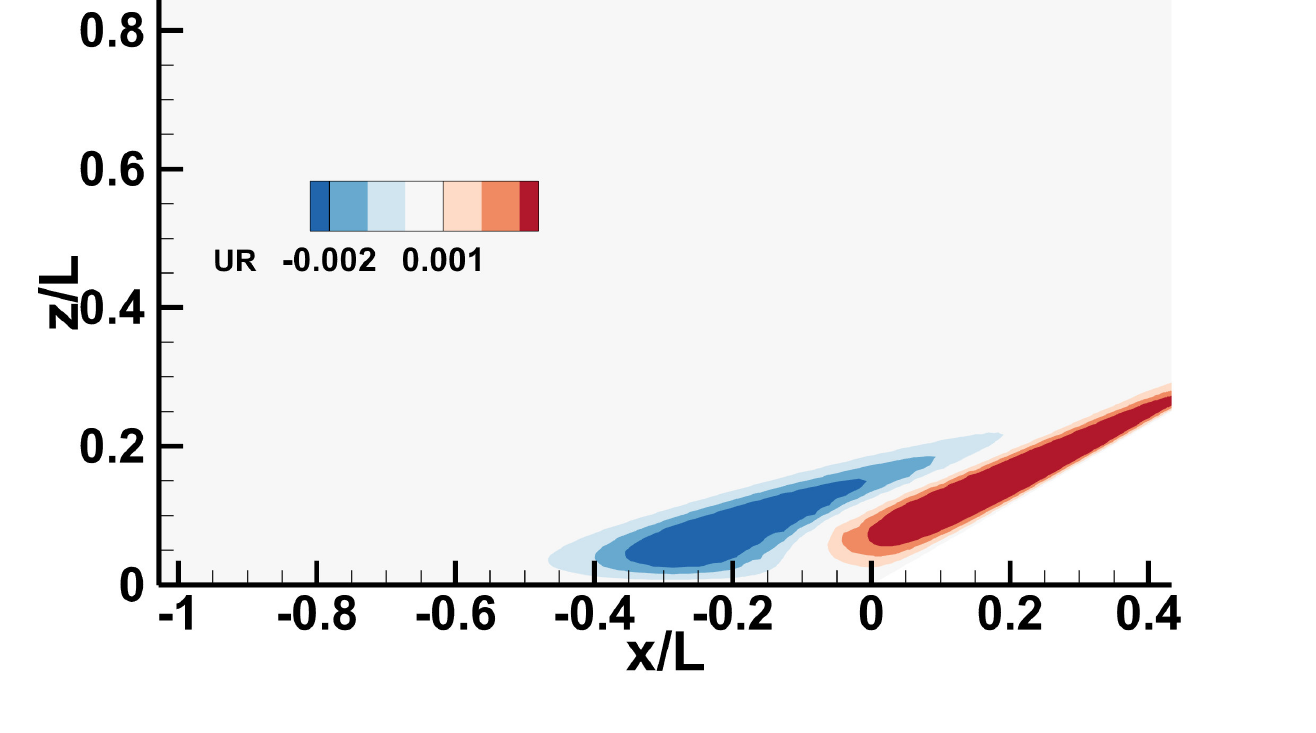}}
\subfigure[]{\label{fig:Mode1sweep30}\includegraphics[trim=70 10 80 10,clip,width=0.48\linewidth]{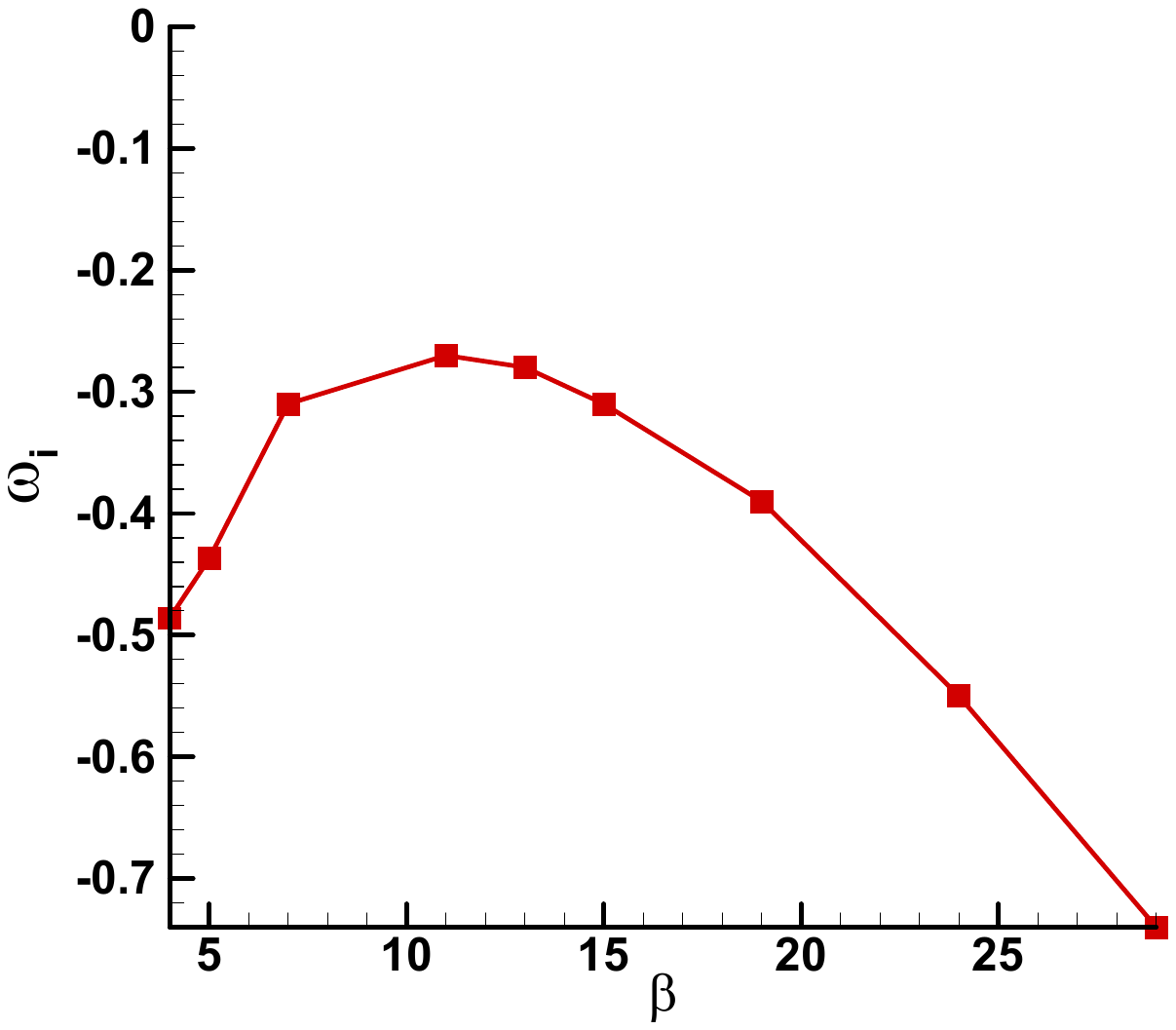}}
\caption{Real part of eigenvectors of spanwise velocity (a) and spanwise wavenumber sweep for the leading mode (b) for 30$^{\circ}$ angle case.}
\label{fig:ModeSweep30}
\end{figure}

\begin{figure}
\center
\subfigure[]{\label{fig:Mode142}\includegraphics[trim=80 10 80 2250,clip,width=0.48\linewidth]{./FIGURES/Fig8a.pdf}}
\subfigure[]{\label{fig:Mode142detail}\includegraphics[trim=70 10 80 250,clip,width=0.48\linewidth]{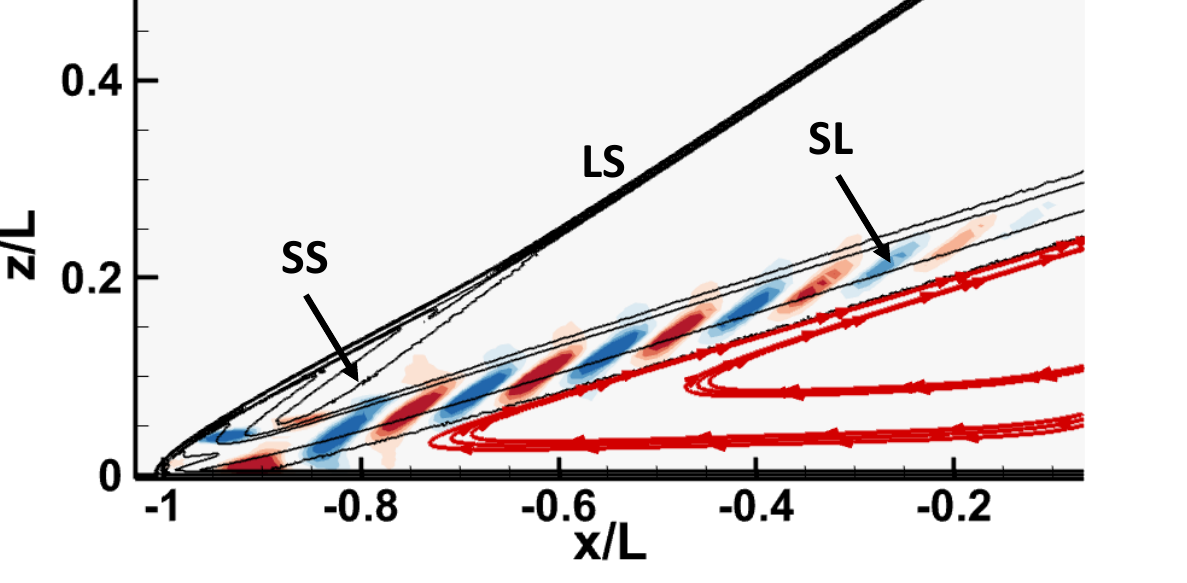}}
\subfigure[]{\label{fig:Mode242}\includegraphics[trim=80 10 80 300,clip,width=0.48\linewidth]{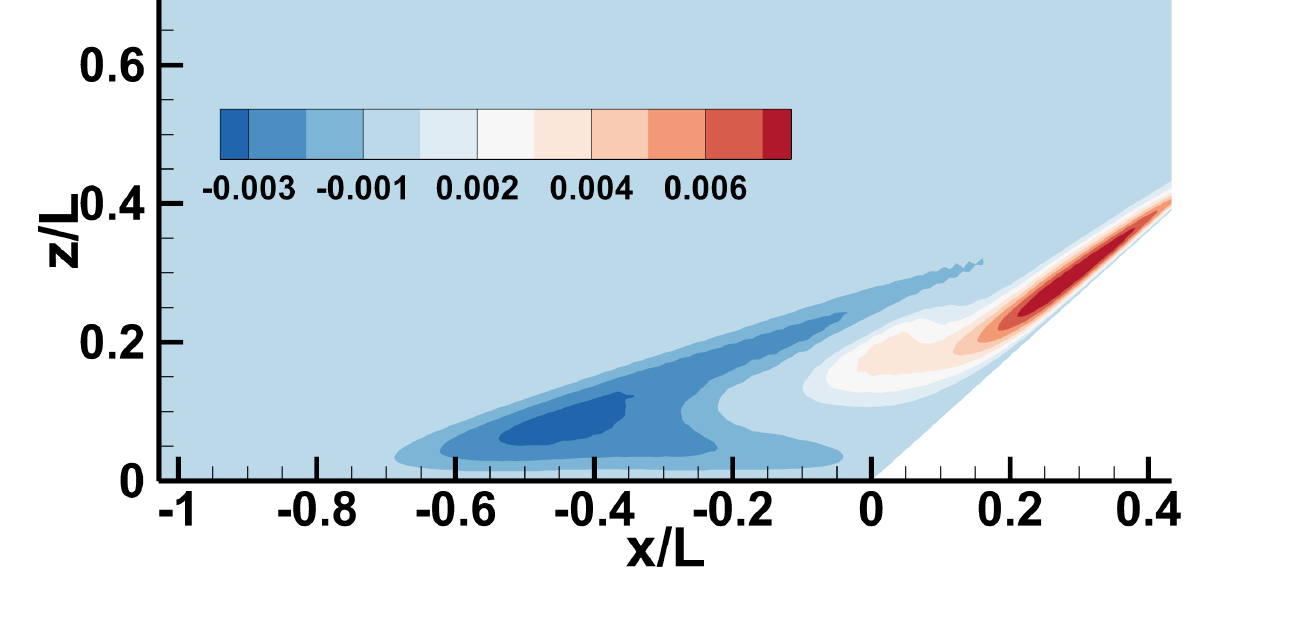}}
\caption{(a): Real part of the amplitude function of the spanwise velocity component, $\hat{w}$, of the leading mode at  42$^{\circ}$, (b) enlarged region near the leading edge and (c) the C-shaped mode also found in the spectrum.}
\label{fig:ModeSweep42}
\end{figure}

\begin{figure}
\center
\includegraphics[trim=70 20 70 20,clip,width=0.48\linewidth]{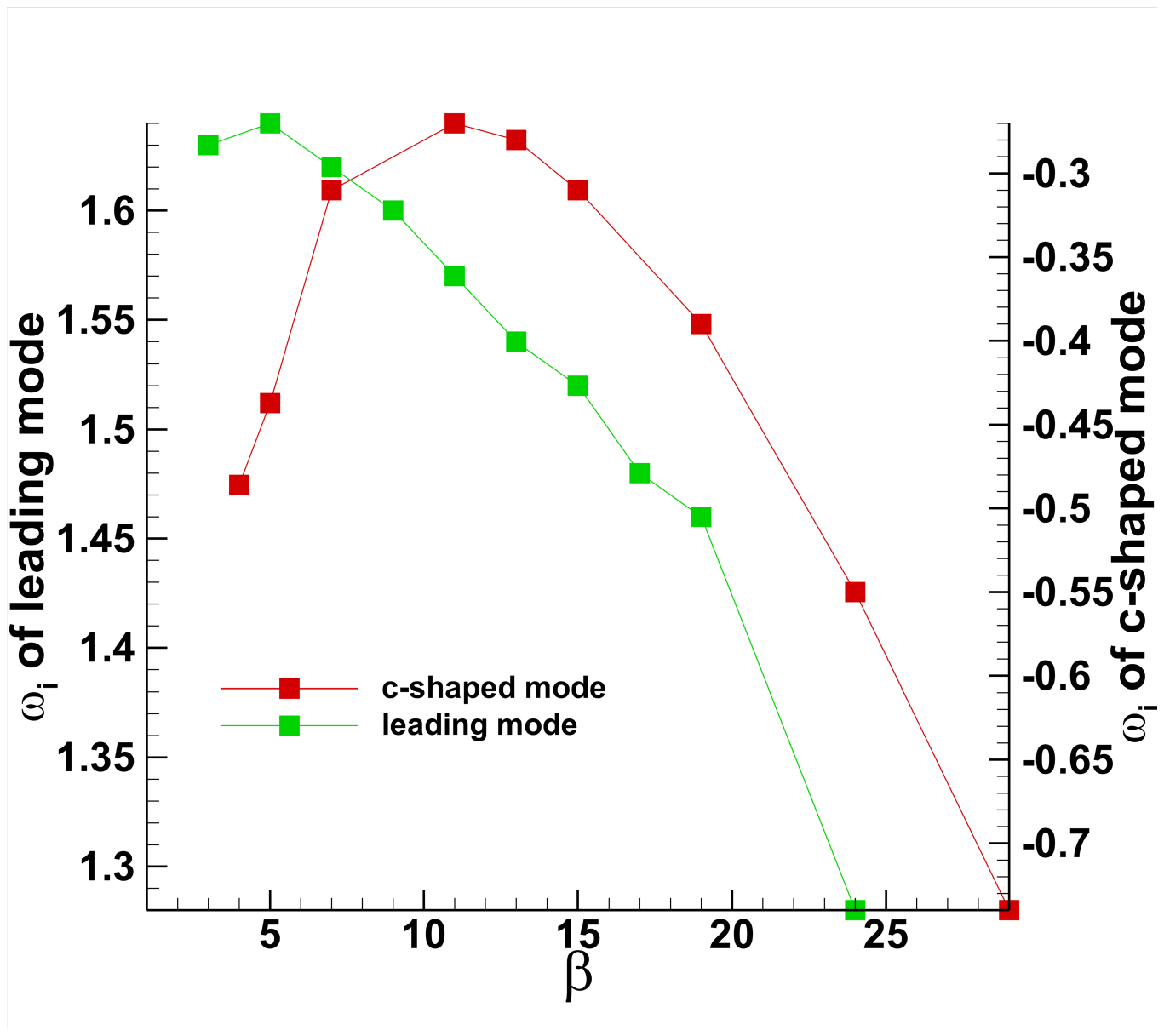}
\caption{Spanwise wavenumber sweep for the leading edge and the C-shaped mode at 42$^{\circ}$.}
\label{fig:ModeSweep42}
\end{figure}

\begin{figure}
\center
\includegraphics[trim=80 10 80 10,clip,width=0.5\linewidth]{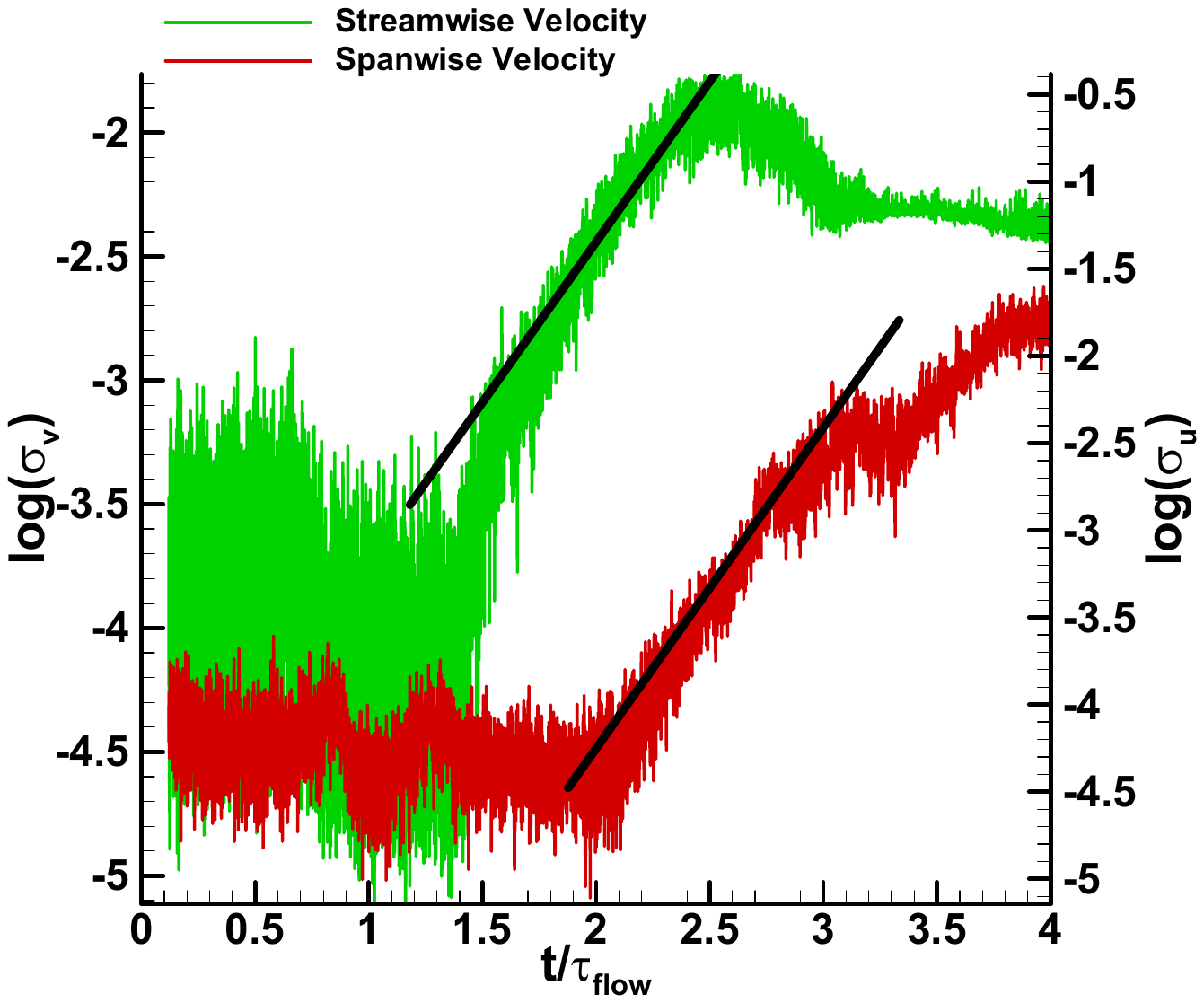}
\caption{Deviations in spanwise and streamwise velocities as a function of normalized time from the 3D-SP DSMC simulations (see Eqn.~\ref{eqn:rms} for the 42$^{\circ}$ ramp.)  DSMC data was taken from the respective spanwise and streamwise probe locations   shown  in Fig.~\ref{fig:IsosurfaceXVorticity}.   Black lines have a  slope corresponding to the growth rate of $\omega_{i}=1.58$ obtained from global LST.}
\label{fig:AmpRates}
\end{figure}

Eigenvalue spectra for both {\color{black} ramp angles} at $\beta=11$ are shown in Fig.~\ref{fig:spectra30}, while Fig.~\ref{fig:spectra42} shows the detail for the stable modes. It can be seen that at 30$^{\circ}$  the leading mode (designated as C-shaped mode, due to the shape of the spanwise perturbation velocity component, see Fig.~\ref{fig:Mode130}) is stationary and no unstable modes are present in the spectrum. A spanwise wavenumber sweep revealed that this mode remains stable, its least damping occurring around $\beta=11$, see Fig.~\ref{fig:Mode1sweep30}.  However, at 42$^{\circ}$  a new unstable traveling mode is discovered, having $\omega=1.58i \pm32 $. The amplitude functions of this mode, identified for the first time here, peak at the region of the interaction of the leading edge shock, separation shock and separated shear layer, see Figs.~\ref{fig:Mode142} and ~\ref{fig:Mode142detail}.  A spanwise wavenumber sweep, shown in Fig.~\ref{fig:ModeSweep42}, illustrated that the leading edge mode is present and always unstable at a wide range of wavenumbers. At these conditions, the C-shaped mode is also present as the second in significance linear global mode, although at this ramp angle it is less damped than at 30$^{\circ}$, see Fig.~\ref{fig:spectra42}.  

\textcolor{black}{Two points are worthy of discussion here. First, the newly discovered traveling mode is absent in the spectrum if the LE shock is excluded from the analysis and the known stationary damped mode dominates the spectrum, as it was shown in the preliminary work of \cite{KarpuzcuCRAviation2022,KarpuzcuCRSciTech2023} and further details can be found in \cite{CerulusPhD}. Second, it is unclear whether the traveling mode {\color{black} would} appear in  spectra obtained using a base state generated by the continuum equations. Our analysis shows that this mode has peaks {\color{black} both at the shear layer, which is well-resolved by the continuum equations, but also inside} the LE shock, the internal structure of which is inaccessible to the Navier-Stokes equations; answers to the latter question require Navier-Stokes base flow simulations that will be reported elsewhere.} 

To further support the present discovery, time-accurate spanwise and streamwise {\color{black} probe data were extracted from the full three-dimensional DSMC} computations, {\color{black} in order to compute from first-principles} the growth rate using the {\color{black} logarithmic derivative of the} deviation of the spanwise and streamwise velocities from their corresponding {\color{black} two-dimensional values}. Fig.~\ref{fig:AmpRates} presents values of  $\sigma_{u}$ and $\sigma_{v}$ computed from,
 \begin{equation}
\label{eqn:rms}
{\sigma _v}(t) = \sqrt {\frac{1}{N}\sum\limits_1^N {{{\left( {\frac{{v(t)}}{{{U_\infty }}}} \right)}^2}} } , \textcolor{white}{0000}{\rm{  }}{\sigma _u}(t) = \sqrt {\frac{1}{N}\sum\limits_1^N {{{\left( {\frac{{u(t) - {u_{2D}}(t)}}{{{U_\infty }}}} \right)}^2}} }
\end{equation}
\textcolor{black}{where $u_{2D}$ is the instantaneous probe data from the 2D DSMC simulations.} Fig.~\ref{fig:AmpRates} shows {\color{black} raw data computed in this manner, alongside the theoretical value computed in the preceding analysis and shown as dark solid lines, the slopes of which match the reported $\omega_i$ value. It can be seen} that the slope of the linear growth of both $\sigma_{u}$ and $\sigma_{v}$ seen {\color{black} in the DSMC simulation} between $t/\tau_{flow}=2-2.5$ {\color{black} matches well that } of the {\color{black} traveling leading-edge mode computed by solution of the eigenvalue problem.  
{\color{black} It is noted that in the evaluation of the DSMC signals, a total of 40 numerical probes were distributed into the flowfield, 20 placed in the spanwise direction near the separation location at $x/L=-0.40$ and 20 in the recirculation region at $x/L=0.10$.} Root mean square values were averaged over the data from these 40 probes, whose locations are shown in Fig.~\ref{fig:IsosurfaceXVorticity}.



\section{Nonlinear simulations at the 42$^{\circ}$ ramp angle}\label{sec:Unsteady42Ramp}

In the previous section it was shown that at 42$^{\circ}$ ramp angle the flow becomes three dimensional \vt{by self-excitation due to linear modal growth of the newly-discovered leading-edge global mode}. Here, the flowfield characteristics of the separation bubble at this ramp angle will be investigated in detail after \vt{unsteadiness and exponential growth of the leading eigenmode has led flow to non-linearity}. To better describe the internal structure of the initially steady and spanwise homogeneous separation bubble, isosurfaces of streamwise vorticity are shown in Fig.~\ref{fig:IsosurfaceXVorticity}.
\vt{ Isosurfaces are plotted at two values of $\omega_x/\tau_{flow}=0.5$ (red) and $-0.5$ (blue); also shown \vt{in greyscale} is the \vt{numerical Schlieren} plane cut at $y/L=1.1$ on which the magnitude of the  number density gradient is plotted.}
The alternating colors of the streamwise vorticity make clear that counter-rotating \vt{streamwise} vortices are \vt{generated at the reattachment region} in the 3D separation bubble. The most salient feature of this flow is the emergence of  $\Lambda$-vortices downstream of the reattachment location, \textcolor{black}{as indicated in the region RF indicated in the figure}. These vortices are \vt{hallmarks of impending transition to turbulence} and have been abundantly observed in \vt{laminar-turbulent transition literature, including experiments, the early temporal \citep{GuoAdamsKleiser1995,AdamsKleiser1996} and spatial \citep{Rist_Fasel_1995,MAYER_WERNZ_FASEL_2011} flat-plate laminar boundary layer Direct Numerical Simulations of transition studies at several Mach numbers, as well as in the recent work of \cite{dwivedi_etal_2022} on the related double wedge configuration}. The seemingly random isosurfaces of the vorticity upstream of the separation line, \textcolor{black}{ indicated by the flow region AF in Fig.~\ref{fig:IsosurfaceXVorticity}}, are due to the expected thermal fluctuations of the DSMC solution as these are instantaneous flowfields, i.e. no time-averaging of the signal has been done.  The structures become organized into larger structures of alternating colors in the separation region, \textcolor{black}{shown in the flow region SR in Fig.~\ref{fig:IsosurfaceXVorticity}}, indicating the counter rotating vortical structures. \vt{To the best of the authors' knowledge, our work is the first in which the early stages of laminar-turbulent transition have been simulated in the context of kinetic theory.}

\begin{figure}
\center
{\includegraphics[trim=80 10 80 10,clip,width=0.75\linewidth]{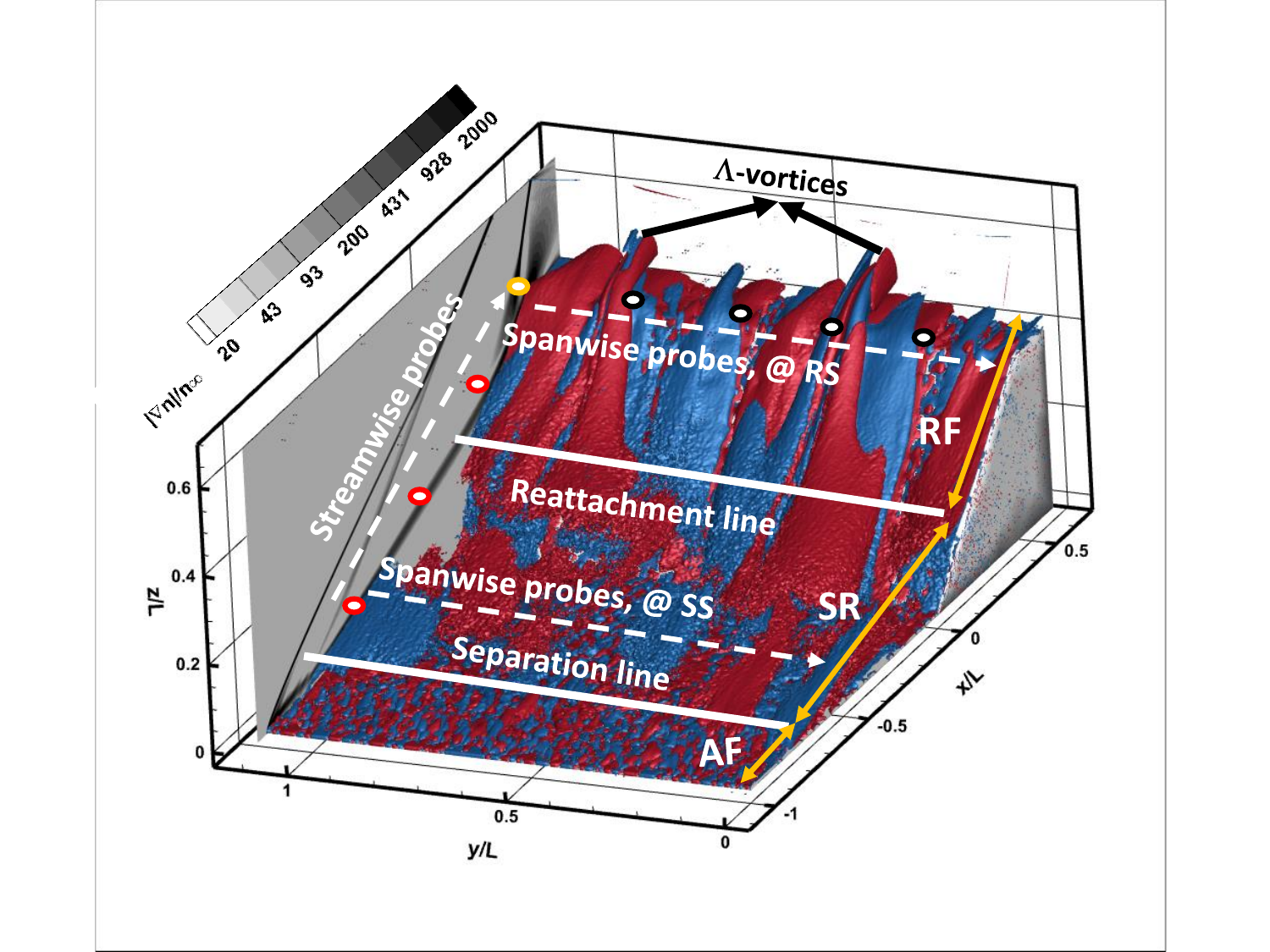}}
\caption{Isosurfaces of instantaneous streamwise vorticity at $t/\tau_{flow}=8$ with the spanwise length of $\lambda=1.12L$ at 42$^{\circ}$ ramp angle. \textcolor{black}{AF: attached flow, SR: separation region, RF: reattached flow.} }
\label{fig:IsosurfaceXVorticity}
\end{figure}

\vt{The ability of DSMC to} resolve the high gradient layers \vt{and, in particular,} rigorously representing the velocity distribution of gas molecules in the \vt{internal structure of a} shock \vt{layer}, rather than treating the shock as a discontinuity, as done in continuum approaches, \vt{is well-established \citep{Bird_1967,Bird_1968,Bird1970,alsmeyer1974messung,Alsmeyer_1976}}.  
  \textcolor{black}{In fact, the shock layer \vt{in the present configuration at the late transitional stages} is not planar and because of its interaction with the shear layer it exhibits the same spanwise periodicity \vt{as that arising from self-excitation of the linear global mode},   \vt{as} seen in the $x-y$ planar cuts of Fig.~\ref{fig:ZPlaneCuts}.  In both Figs.~\ref{fig:Zplanedown} and ~\ref{fig:Zplaneup}, it can be observed that the leading edge shock (LS), \ is almost uniform in the spanwise direction prior to its interaction with the separation shock (SS)\vt{. This spanwise uniformity is lost} 
    after these two shocks merge, \vt{owing to the } interaction \vt{of the separation shock with the separated shear layer (SL).}} After the leading edge and separation shocks merge \textcolor{black}{near $x/L=-0.1$ in Fig.~\ref{fig:Zplanedown} and  $x/L=0.1$ in Fig.~\ref{fig:Zplaneup}}, the combined structure becomes spanwise varying. Moreover, as the reattachment shock (RS) forms, it also has the same spanwise periodic shape as the shear layer due to its interaction with the  $\Lambda$-vortices. These observations agree with the previous work of ~\cite{sawant_etal_2022} where it was shown that the instabilities inside the shock layer synchronize with \vt{ those of the laminar separation bubble formed at that compression corner and its} shear layer.

\begin{figure}
\center
\subfigure[]{\label{fig:Zplanedown}\includegraphics[trim=80 10 100 10,clip,width=0.48\linewidth]{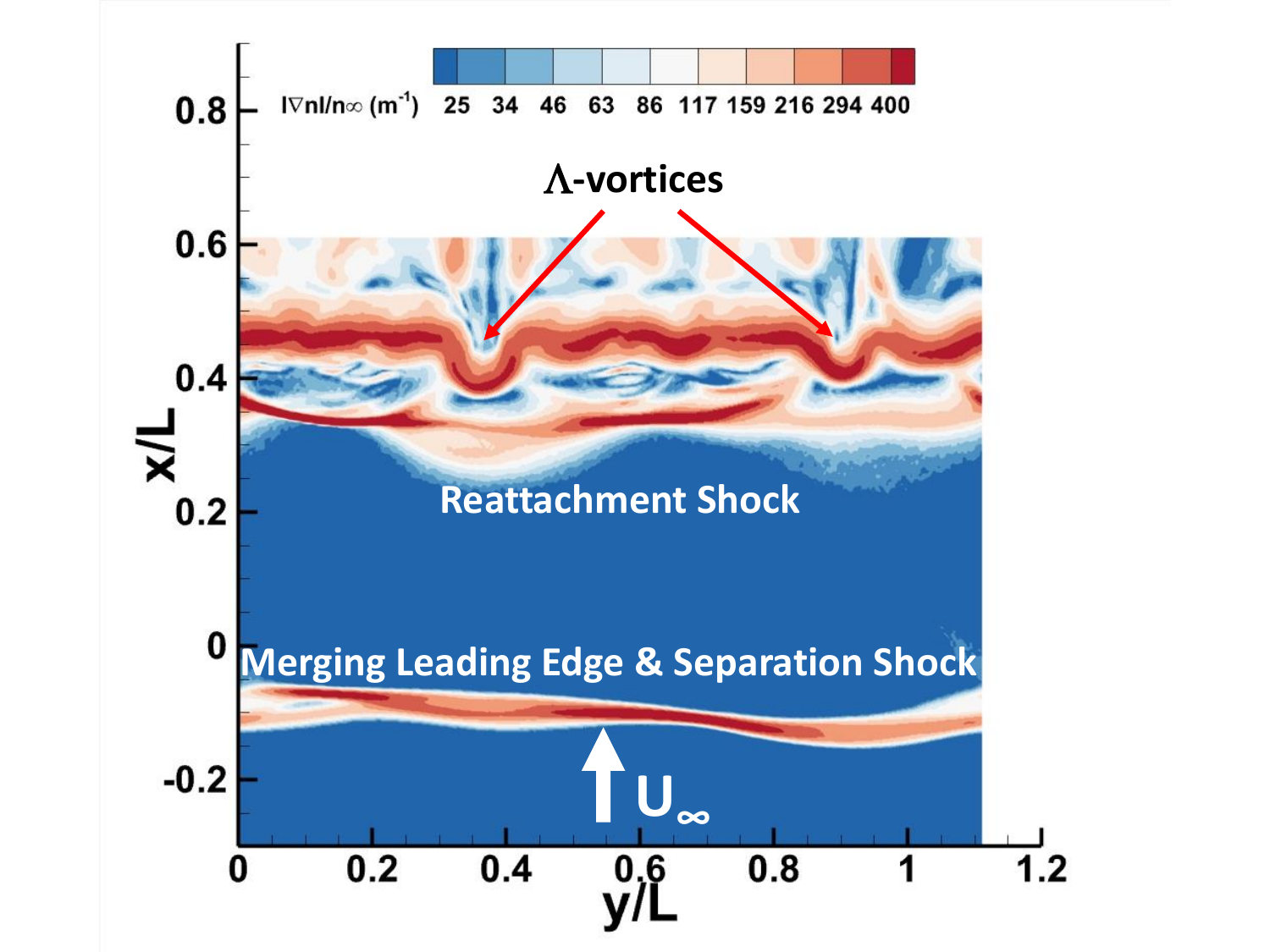}}
\subfigure[]{\label{fig:Zplaneup}\includegraphics[trim=80 10 100 10,clip,width=0.48\linewidth]{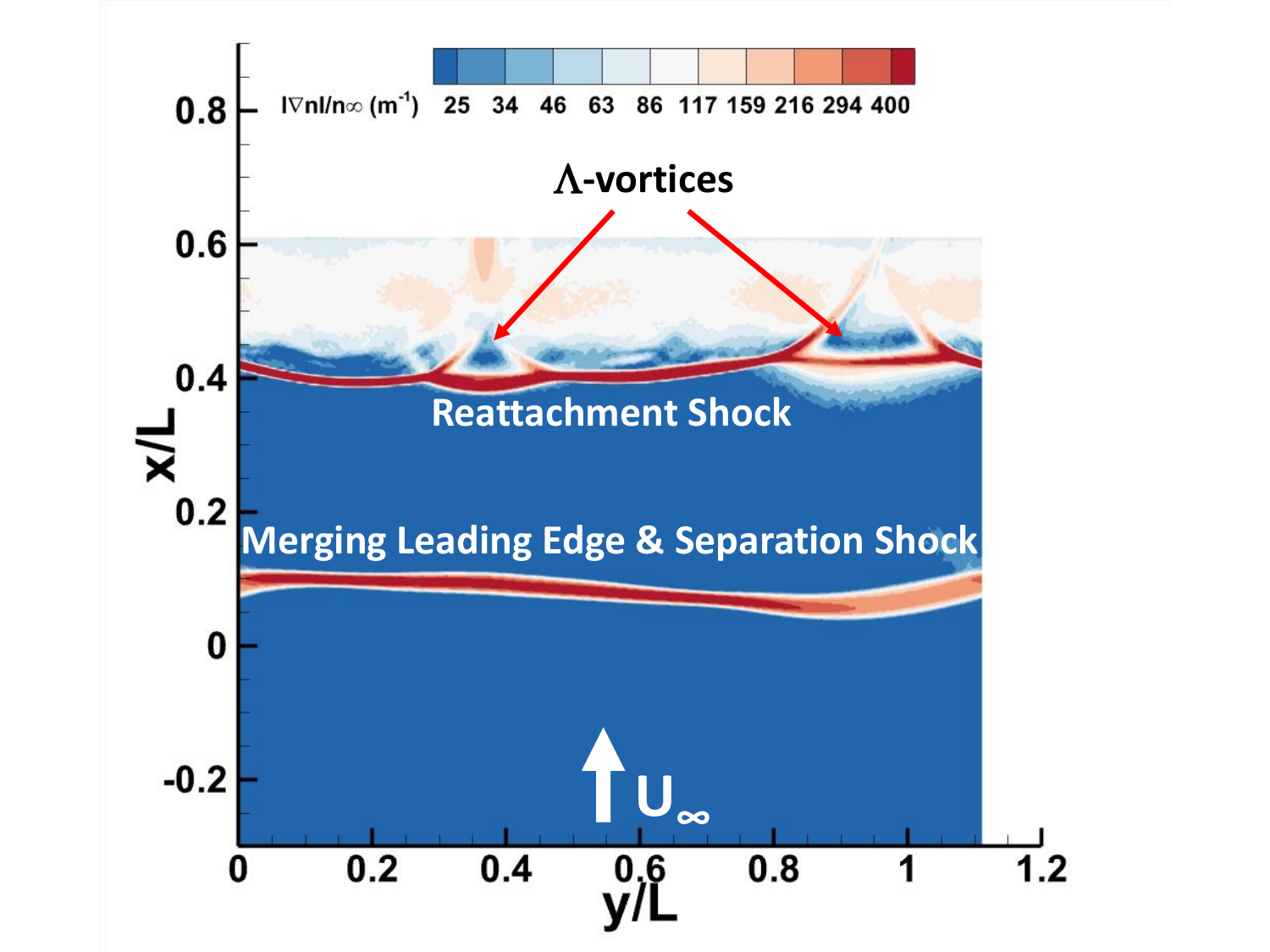}}
\caption{x-y plane cuts showing the wall normal evolution of the shock and streamwise velocity at (a) $z/L=0.44$ and (b) $z/L=0.56$. Flooded contours show magnitude of the gradient of number density normalized with the free stream number density.}
\label{fig:ZPlaneCuts}
\end{figure}

\begin{figure}
\center
\subfigure[]{\label{fig:PressReattachProbes}\includegraphics[trim=80 10 80 10,clip,width=0.48\linewidth]{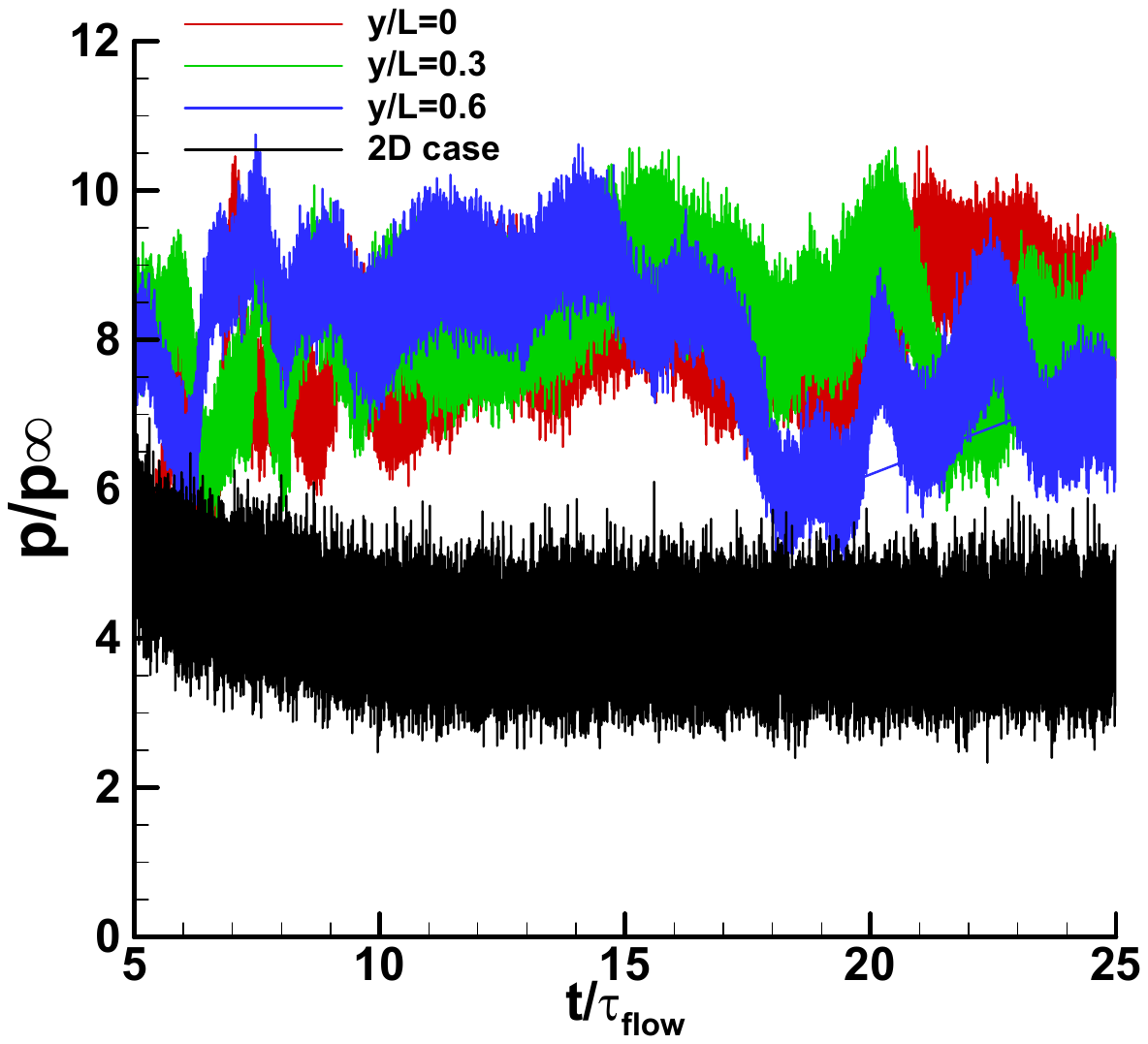}}
\subfigure[]{\label{fig:PressReattachPSD}\includegraphics[trim=70 10 70 10,clip,width=0.48\linewidth]{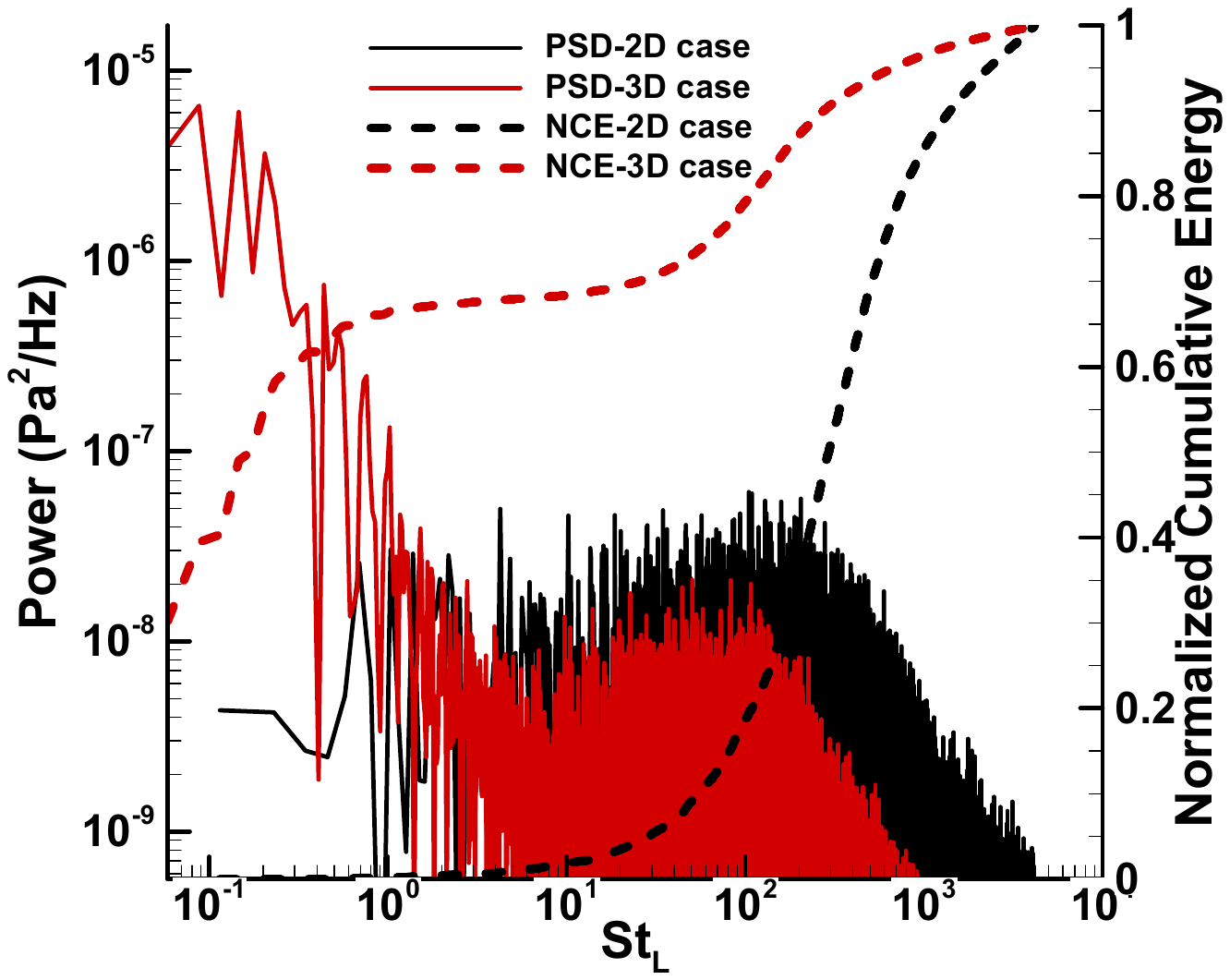}}
\caption{(a) Time evolution of pressure, comparing 2D results with those at different 3D spanwise locations. (b) PSD of pressure signal at probe locations near reattachment region at $x/L=0.38$ and $z/L=0.37$. PSD for the 3D case corresponds to the probe at $x/L=0.38$, $y/L=0$ and $z/L=0.37$. }
\label{fig:Probe4data}
\end{figure}

In addition to the 3D structures and spanwise variation in the flowfield velocities, it was observed that the \vt{three-dimensional} flow is highly unsteady, particularly near the separation and reattachment shocks.
Non-linear oscillations are observed after the initial linear growth period, between $t/\tau_{flow}=2-2.5$, as shown in Fig.~\ref{fig:PressReattachProbes}, and the flow becomes three dimensional and unsteady, \textcolor{black}{as already seen in Figs.~\ref{fig:IsosurfaceXVorticity} and \ref{fig:ZPlaneCuts}}.  In Fig.~\ref{fig:PressReattachProbes} the temporal evolution of pressure is shown at times after $t/\tau_{flow}=5$, {\color{black} at a location} near the reattachment shock region, $x/L=0.38$ and $z/L=0.37$, in 2D (black curve), {\color{black} as well as on several probes placed along the spanwise direction in 3D simulation at the same $(x/L,z/L)$ location}. It can be clearly seen that after five flow times the {\color{black} two-dimensional flow} reaches a steady state whereas all the numerical probes from the {\color{black} in the three-dimensional simulation exhibit} unsteadiness. For spectral estimation of the instantaneous DSMC data, the Fast Fourier Transform (FFT) algorithm is used and Power Spectral Densities (PSD) have been calculated by multiplying FFTs in each frequency bin with their complex conjugate. PSD analysis of the pressure values from the DSMC numerical probes \vt{shown in } Fig.~\ref{fig:PressReattachPSD} also \vt{verifies{}} that the \vt{strictly two-dimensional flow} is steady, since no {\color{black} significant} peaks are observed  \vt{beyond} the peaks due to the thermal fluctuations present in the DSMC data. In contrast, the pressure data from the \vt{corresponding three-dimensional simulation} shows  peaks in the low frequency region centered around a Strouhal number $S{t_L} = fL/{U_\infty }=0.15$. {\color{black} This} is confirmed by the inflection point in the normalized cumulative energy (NCE) curves, calculated by adding up the power in each frequency bin of $70 Hz$ of the PSD  analysis as given by
\begin{equation}
\label{eqn:NCE}
CE(F) = \sum\limits_{f = 0}^{f = F} {PSD(f)}
\end{equation}
and self normalizing with the maximum cumulative energy.

Furthermore, the synchronization of flow oscillations among probes along the streamwise direction was \vt{quantified by} the cross correlation of streamwise velocity. \vt{In order to perform} comparisons, probes \vt{are placed along several locations on two planes, using as an anchor point a location near the reattachment shock, indicated by an orange dot in} Fig.~\ref{fig:IsosurfaceXVorticity}, \vt{and having coordinates}  ($x/L, y/L, z/L =0.38,1.09, 0.37$).  \vt{Three probes indicated by black dots are placed on}  the same y-z plane as the base (orange) probe and are  used to obtain the  spanwise correlation. \vt{Analogously,} three probes indicated by red dots \vt{are placed on} the same x-z plane as the base (orange) probe are considered  \vt{to compute} streamwise correlations. The  cross correlation relation is \vt{defined by},
 \begin{equation}
\label{eqn:CrossCorr}
{R_{AB}}(m) = \frac{{\sum\limits_{t = 0}^{T - m - 1} {{A_{t + m}}{B_t}} }}{{\sqrt {{R_{AA}}(0){R_{BB}}(0)} }},
\end{equation}
where $A$ and $B$ are time series data from two different probe points, $T$ is the size of the vector, $t$ is \vt{DSMC data averaged over 0.05~ms and considered as "instantaneous"}, and $m$ is the off-set or "time lag" between the two time series. Note that $A$ is always the time series data from the base probe, shown with the orange dot in Fig.~\ref{fig:IsosurfaceXVorticity}, and $B$ represents time series data from every other streamwise and spanwise probe, thus, the cross correlation of the base probe is calculated for each of the probes in those directions. The correlation is normalized by the auto-correlation of the time series data with themselves at zero lag.

 The time evolution of the streamwise velocity \vt{at} the streamwise probes is shown in Fig.~\ref{fig:VxStreamProbes} after the flow becomes fully oscillatory and non-linear.  \textcolor{black}{Even without additional analysis, the  oscillations in the DSMC data for these streamwise velocities from the four different probes can be already identified as synchronized.} To ensure that DSMC thermal fluctuations do not affect cross correlation calculations, time-resolved data is averaged every 1,000 time steps, corresponding to $0.05~ms$.

The output of the cross-correlation equation (\ref{eqn:CrossCorr}) provides two variables, the correlation coefficient, $R$, which represents how similar the two time-series data is and the time offset, $m$, which represents time lag. For each value of the time off-set, $m$, a correlation coefficient, $R$, is calculated.  Fig.~\ref{fig:VxCrossCorr} shows the corresponding cross correlation values for streamwise (red) and spanwise (black) correlations. It can be seen that the oscillations of the streamwise velocity in the streamwise direction are highly correlated, having $R>0.96$, unity indicating perfect correlation. Moreover, high correlation coefficient values are found near zero time offset which indicates the synchronization of the signal along the streamwise direction. By contrast, as seen in the same figure, the spanwise variation of the velocity signal has much smaller correlation values, which also exhibit no peaks near the zero lag point. Both results \textcolor{black}{further confirm the streamwise coherence of the streamwise aligned structures shown in Fig.~\ref{fig:IsosurfaceXVorticity}.}


\begin{figure}
\center
\subfigure[]{\label{fig:VxStreamProbes}\includegraphics[trim=80 10 80 10,clip,width=0.48\linewidth]{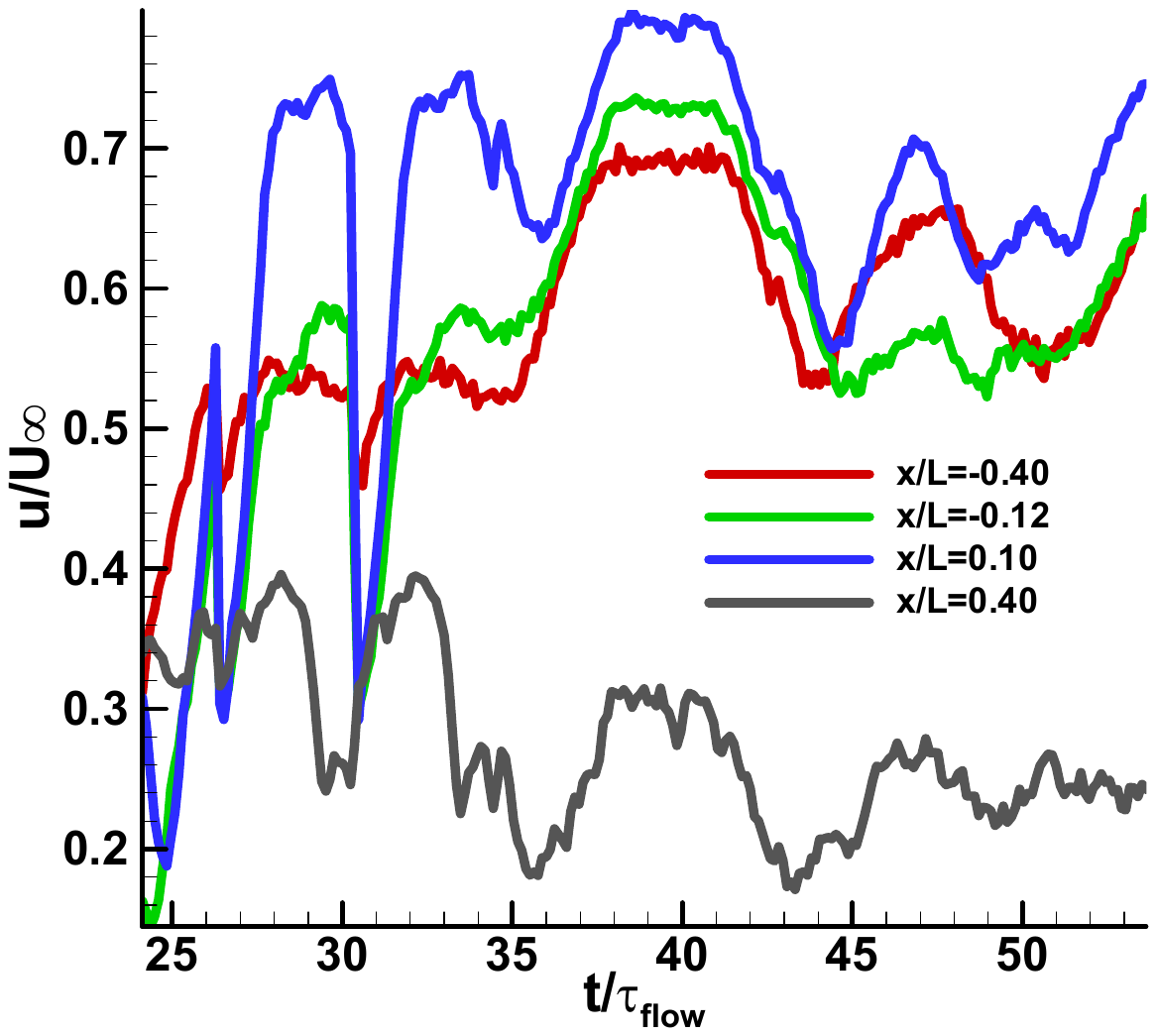}}
\subfigure[]{\label{fig:VxCrossCorr}\includegraphics[trim=80 10 80 10,clip,width=0.48\linewidth]{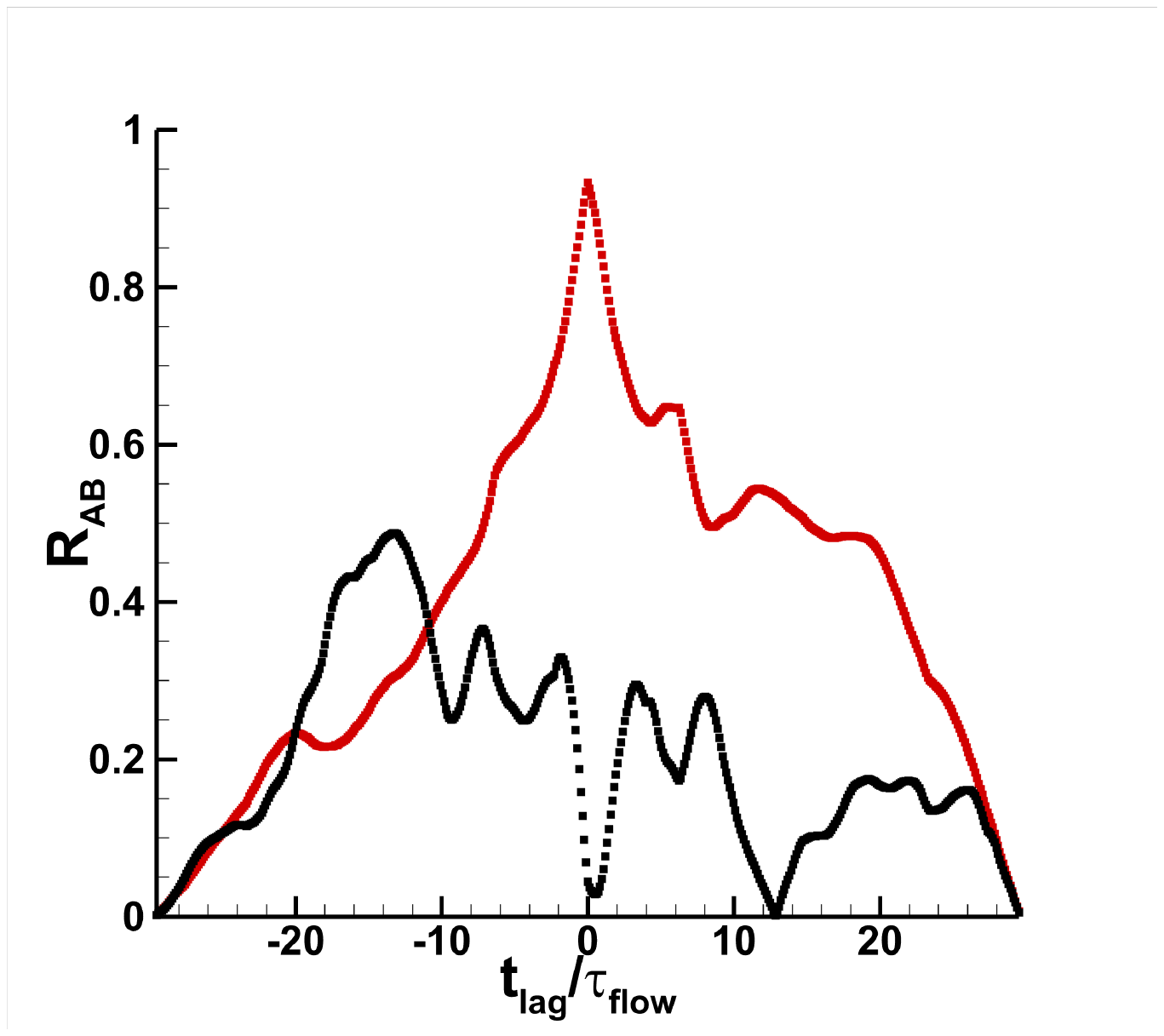}}
\caption{Time evolution of the streamwise velocity along streamwise direction (a) and cross correlation of the streamwise velocity along streamwise (red curve) and spanwise (black curve) directions (b). Probe locations are shown in Fig.~\ref{fig:IsosurfaceXVorticity}.}
\label{fig:CrossCorr}
\end{figure}

\pagebreak
\section{Conclusions}
\label{sec:conclusions}
Mach 3 flow over \vt{two short, planar} compression corners \vt{was analyzed  numerically using two- and three-dimensional DSMC simulations and global linear stability theory. A low Reynolds number was chosen, sufficiently large for the continuum assumption to be valid,}
and two relatively large ramp angles were considered. \vt{Two-dimensional} DSMC \vt{simulations at both angles result in steady large laminar separation bubbles, featuring a single recirculation zone which practically extends over the entire length of the plate. Despite the relatively large scaled angle values, at which two-dimensional flow would be expected to be unsteady according to the criteria put forward by triple-deck theory \citep{cassel_ruban_walker_1995,egorov2011three}, at both ramp angles steady two-dimensional flows were computed; the essentially different from continuum theory relatively large slip velocities computed in the DSMC may account for this qualitative difference.} \vt{Another intriguing aspect of the steady laminar two-dimensional base flows is the relatively large, more than $10\%$, maximum value of the reverse velocity within the separation bubbles at both angles, which could have led to the expectation of self-excitation of the stationary spanwise-periodic global mode of laminar separation bubbles \citep{Theofilis2000,BoinEtAlTCFD}}.

\vt{However, three-dimensional }
modal BiGlobal stability analysis \vt{of} the steady, \vt{laminar, two-dimensional} base flows, \vt{in which, by contrast to all analyses available in the literature, the fully-resolved leading edge shock has been included}, \vt{has delivered a number of interesting results}.  \vt{The} 30$^{\circ}$ ramp angle \vt{configuration} was found to only \vt{support} stable \vt{eigen}modes, the leading of which is the well known \vt{from earlier analyses stationary, three-dimensional C-shaped mode} of the \vt{laminar} separation bubble. \vt{The known mode is also present but also damped} in the 42$^{\circ}$  ramp angle \vt{configuration, however at this higher angle a previously unknown global eigenmode has been discovered, the amplitude functions of which were found to peak at the leading edge shock and along the separation bubble shear layer.} A spanwise wavenumber sweep showed that this \vt{ Leading Edge (LE) mode} is unstable at a wide range of wavenumbers. Three-dimensional, \vt{(unsteady)} spanwise periodic DSMC \vt{simulations} confirmed the LST predictions \vt{in both configurations, delivering} no spanwise variation \vt{at} 30$^{\circ}$ \vt{and spanwise-periodic structures at} 42$^{\circ}$ \vt{with} spanwise length \vt{that} of the unstable \vt{global} mode. \vt{The exponential growth of this mode, as predicted from post-processing of} time accurate data  \vt{obtained at} numerical probes \vt{placed in the flow field of DSMC simulations}, agrees very well with the growth rate predicted by the \vt{linear theory}. This result is analogous to that obtained in the recent DSMC-based work of \cite{sawant_etal_2022} and again demonstrates the \vt{ability of} kinetic simulations \vt{to perform} linear stability \vt{analysis and deliver results inaccessible to continuum-based analysis, when instability originates in the internal structure of the shock layer}. \vt{Finally, the late stages of transition were probed in the present DSMC simulations when the leading linear perturbation grew to nonlinear levels and} $\Lambda$-vortices, known to appear ahead of transition to turbulence, were observed to form downstream of flow reattachment. Oscillations present in the flow velocities \vt{at this stage were found to be} highly correlated along the streamwise direction and poorly correlated along the spanwise direction, indicating \vt{the streamwise alignment of nonlinearly generated structures prior to transition. To the best of the authors' knowledge, this is the first time that the late laminar-turbulent transition stages have been probed by kinetic theory methods in any high-speed configuration.}

\section*{Acknowledgments}
The research conducted in this paper is supported by the Office of Naval Research under Grant No. N000141202195 titled “Multi-scale modeling of unsteady shock-boundary layer hypersonic ﬂow instabilities,” with Dr. Eric Marineau as the Program Officer. This research is also supported by NSF ACCESS (previously XSEDE) Frontera supercomputer with project numbers CTS22009 and CTS23002. The authors are grateful to Drs. Helio Quintanilha \vt{and Nic Cerulus} for providing the LIGHT code \vt{set up for compression ramp analyses, as well as for} useful discussions.

\section*{Declaration of Interests}
The authors report no conflict of interest

\bibliography{SpanwiseRamp}

\begin{thebibliography}{80}
\expandafter\ifx\csname natexlab\endcsname\relax\def\natexlab#1{#1}\fi
\def\au#1{#1} \def\ed#1{#1} \def\yr#1{#1}\def\at#1{#1}\def\jt#1{\textit{#1}}
  \def\bt#1{#1}\def\bvol#1{\textbf{#1}} \def\vol#1{#1} \def\pg#1{#1}
  \def\publ#1{#1}\def\arxiv#1{#1}\def\org#1{#1}\def\st#1{\textit{#1}}

\bibitem[Adams \& Kleiser(1996)]{AdamsKleiser1996}
{\sc \au{Adams, N.~A.} \& \au{Kleiser, L.}} \yr{1996}  \at{Subharmonic
  transition to turbulence in a flat-plate boundary layer at {M}ach number
  4.5}.  \jt{Journal of Fluid Mechanics}  \bvol{317},  \pg{301–335}.

\bibitem[Alsmeyer(1974)]{alsmeyer1974messung}
{\sc \au{Alsmeyer, H.}} \yr{1974}  \at{Messung der struktur von
  verdichtungsst{\"o}ssen in argon und stickstoff}. PhD thesis, University of
  Karlsruhe, Germany.

\bibitem[Alsmeyer(1976)]{Alsmeyer_1976}
{\sc \au{Alsmeyer, H.}} \yr{1976}  \at{Density profiles in argon and nitrogen
  shock waves measured by the absorption of an electron beam}.  \jt{Journal of
  Fluid Mechanics}  \bvol{74}~(3),  \pg{497–513}.

\bibitem[Andreopoulos \& Muck(1987)]{Andreopoulos_Muck_1987}
{\sc \au{Andreopoulos, J.} \& \au{Muck, K.~C.}} \yr{1987}  \at{Some new aspects
  of the shock-wave/boundary-layer interaction in compression-ramp flows}.
  \jt{Journal of Fluid Mechanics}  \bvol{180},  \pg{405–428}.

\bibitem[Benitez {\em et~al.\/}(2023)Benitez, Borg, Scholten, Paredes, McDaniel
  \& Jewell]{benitez_borg_scholten_paredes_mcdaniel_jewell_2023}
{\sc \au{Benitez, E.~K.}, \au{Borg, M.~P.}, \au{Scholten, A.}, \au{Paredes,
  P.}, \au{McDaniel, Z.} \& \au{Jewell, J.~S.}} \yr{2023}  \at{Instability and
  transition onset downstream of a laminar separation bubble at {M}ach 6}.
  \jt{Journal of Fluid Mechanics}  \bvol{969},  \pg{A11}.

\bibitem[Bird(1967)]{Bird_1967}
{\sc \au{Bird, G.~A.}} \yr{1967}  \at{The velocity distribution function within
  a shock wave}.  \jt{Journal of Fluid Mechanics}  \bvol{30}~(3),
  \pg{479–487}.

\bibitem[Bird(1968)]{Bird_1968}
{\sc \au{Bird, G.~A.}} \yr{1968}  \at{The structure of normal shock waves in a
  binary gas mixture}.  \jt{Journal of Fluid Mechanics}  \bvol{31}~(4),
  \pg{657–668}.

\bibitem[Bird(1970)]{Bird1970}
{\sc \au{Bird, G.~A.}} \yr{1970}  \at{{Aspects of the Structure of Strong Shock
  Waves}}.  \jt{The Physics of Fluids}  \bvol{13}~(5),  \pg{1172--1177}.

\bibitem[Bird(1994)]{Bird}
{\sc \au{Bird, G.~A.}} \at{ \yr{1994} } \bt{In {\em Molecular Gas Dynamics and
  the Direct Simulation of Gas Flows\/}}.  \publ{Oxford, England, U.K.:
  Clarendon}.

\bibitem[Bloy \& Georgeff(1974)]{Bloy_Georgeff_1974}
{\sc \au{Bloy, A.~W.} \& \au{Georgeff, M.~P.}} \yr{1974}  \at{The hypersonic
  laminar boundary layer near sharp compression and expansion corners}.
  \jt{Journal of Fluid Mechanics}  \bvol{63}~(3),  \pg{431–447}.

\bibitem[Boin {\em et~al.\/}(2006)Boin, Robinet, Corre \& Deniau]{BoinEtAlTCFD}
{\sc \au{Boin, J.}, \au{Robinet, J.~C.}, \au{Corre, C.} \& \au{Deniau, H.}}
  \yr{2006}  \at{3{D} steady and unsteady bifurcations in a shock-wave/laminar
  boundary layer interaction: A numerical study}.  \jt{Theoretical and
  Computational Fluid Dynamics}  \bvol{20}~(3),  \pg{163--180}.

\bibitem[Cao {\em et~al.\/}(2023)Cao, Hao, Guo, Wen \&
  Klioutchnikov]{Cao_Hao_Guo_Wen_Klioutchnikov_2023}
{\sc \au{Cao, S.}, \au{Hao, J.}, \au{Guo, P.}, \au{Wen, C.-Y.} \&
  \au{Klioutchnikov, I.}} \yr{2023}  \at{Stability of hypersonic flow over a
  curved compression ramp}.  \jt{Journal of Fluid Mechanics}  \bvol{957},
  \pg{A8}.

\bibitem[Cao {\em et~al.\/}(2021{\natexlab{{\em a\/}}})Cao, Hao, Klioutchnikov,
  Olivier, Heufer \& Wen]{Cao_Hao_Klioutchnikov_Olivier_Heufer_Wen_2021}
{\sc \au{Cao, S.}, \au{Hao, J.}, \au{Klioutchnikov, I.}, \au{Olivier, H.},
  \au{Heufer, K.~A.} \& \au{Wen, C.-Y.}} \yr{2021{\natexlab{{\em a\/}}}}
  \at{Leading-edge bluntness effects on hypersonic three-dimensional flows over
  a compression ramp}.  \jt{Journal of Fluid Mechanics}  \bvol{923},  \pg{A27}.

\bibitem[Cao {\em et~al.\/}(2021{\natexlab{{\em b\/}}})Cao, Hao, Klioutchnikov,
  Olivier \& Wen]{Cao_Hao_Klioutchnikov_Olivier_Wen_2021}
{\sc \au{Cao, S.}, \au{Hao, J.}, \au{Klioutchnikov, I.}, \au{Olivier, H.} \&
  \au{Wen, C.-Y.}} \yr{2021{\natexlab{{\em b\/}}}}  \at{Unsteady effects in a
  hypersonic compression ramp flow with laminar separation}.  \jt{Journal of
  Fluid Mechanics}  \bvol{912},  \pg{A3}.

\bibitem[Cao {\em et~al.\/}(2022)Cao, Hao, Klioutchnikov, Wen, Olivier \&
  Heufer]{cao_etal_2022}
{\sc \au{Cao, S.}, \au{Hao, J.}, \au{Klioutchnikov, I.}, \au{Wen, C.-Y.},
  \au{Olivier, H.} \& \au{Heufer, K.~A.}} \yr{2022}  \at{Transition to
  turbulence in hypersonic flow over a compression ramp due to intrinsic
  instability}.  \jt{Journal of Fluid Mechanics}  \bvol{941},  \pg{A8}.

\bibitem[Cassel {\em et~al.\/}(1995)Cassel, Ruban \&
  Walker]{cassel_ruban_walker_1995}
{\sc \au{Cassel, K.~W.}, \au{Ruban, A.~I.} \& \au{Walker, J. D.~A.}} \yr{1995}
  \at{An instability in supersonic boundary-layer flow over a compression
  ramp}.  \jt{Journal of Fluid Mechanics}  \bvol{300},  \pg{265–285}.

\bibitem[Cerulus(2022)]{CerulusPhD}
{\sc \au{Cerulus, N.}} \yr{2022}  \at{Characterisation of the stability of
  compression corner geometries under supersonic flow conditions}. PhD thesis,
  University of Liverpool.

\bibitem[Chapman {\em et~al.\/}(1958)Chapman, Kuehn \&
  Larson]{chapman1958investigation}
{\sc \au{Chapman, D.~R.}, \au{Kuehn, D.~M.} \& \au{Larson, H.~K.}} \yr{1958}
  \bt{Investigation of separated flows in supersonic and subsonic streams with
  emphasis on the effect of transition}. {\em Tech. Rep.\/} NACA-TR-1356.
  \org{National Advisory Comittee for Aeronautics}.

\bibitem[Clemens \& Narayanaswamy(2014)]{ClemensReviewSWBLI}
{\sc \au{Clemens, N.~T.} \& \au{Narayanaswamy, V.}} \yr{2014}
  \at{Low-frequency unsteadiness of shock wave/turbulent boundary layer
  interactions}.  \jt{Annual Review of Fluid Mechanics}  \bvol{46}~(1),
  \pg{469--492}.

\bibitem[Cowley \& Hall(1990)]{CowleyHall1990}
{\sc \au{Cowley, S.} \& \au{Hall, P.}} \yr{1990}  \at{On the instability of
  hypersonic flow past a wedge}.  \jt{Journal of Fluid Mechanics}  \bvol{214},
  \pg{17–42}.

\bibitem[Davami {\em et~al.\/}(2024)Davami, Juliano, Scholten, Paredes,
  Benitez, Running, Dylewicz, Pezlar, Theofilis, Thiele \&
  Willems]{DavamiEtAlAIAA2024-0499}
{\sc \au{Davami, J.}, \au{Juliano, T.~J.}, \au{Scholten, A.}, \au{Paredes, P.},
  \au{Benitez, E.~K.}, \au{Running, C.~L.}, \au{Dylewicz, K.}, \au{Pezlar, V.},
  \au{Theofilis, V.}, \au{Thiele, T.} \& \au{Willems, S.}} \yr{2024}
  \at{Separation and transition on the rotex-t cone-flare}.  \jt{AIAA Paper
  2024-0499} .

\bibitem[De~Tullio {\em et~al.\/}(2013)De~Tullio, Paredes, Sandham \&
  Theofilis]{DeTullioEtAlJFM2013}
{\sc \au{De~Tullio, N.}, \au{Paredes, P.}, \au{Sandham, N.~D.} \&
  \au{Theofilis, V.}} \yr{2013}  \at{Laminar–turbulent transition induced by
  a discrete roughness element in a supersonic boundary layer}.  \jt{Journal of
  Fluid Mechanics}  \bvol{735},  \pg{613–646}.

\bibitem[Dolling(2001)]{DollingreviewSWBLI}
{\sc \au{Dolling, D.~S.}} \yr{2001}  \at{Fifty years of
  shock-wave/boundary-layer interaction research: What next?}  \jt{AIAA
  Journal}  \bvol{39}~(8),  \pg{1517--1531}.

\bibitem[Dwivedi {\em et~al.\/}(2022)Dwivedi, Sidharth \&
  Jovanović]{dwivedi_etal_2022}
{\sc \au{Dwivedi, A.}, \au{Sidharth, G.} \& \au{Jovanović, M.~R.}} \yr{2022}
  \at{Oblique transition in hypersonic double-wedge flow}.  \jt{Journal of
  Fluid Mechanics}  \bvol{948},  \pg{A37}.

\bibitem[Dwivedi {\em et~al.\/}(2019)Dwivedi, Sidharth, Nichols, Candler \&
  Jovanović]{dwivedi_etal_2019}
{\sc \au{Dwivedi, A.}, \au{Sidharth, G.~S.}, \au{Nichols, J.~W.}, \au{Candler,
  G.~V.} \& \au{Jovanović, M.~R.}} \yr{2019}  \at{Reattachment streaks in
  hypersonic compression ramp flow: an input–output analysis}.  \jt{Journal
  of Fluid Mechanics}  \bvol{880},  \pg{113–135}.

\bibitem[Edney(1968)]{Edney}
{\sc \au{Edney, B.}} \yr{1968}  \bt{Anomalous heat transfer and pressure
  distributions on blunt bodies at hypersonic speeds in the presence of an
  impinging shock.}  \org{The {A}eronautical {R}esearch {I}nstıtute of
  {S}weden, {R}eport 115}.

\bibitem[Egorov {\em et~al.\/}(2011)Egorov, Neiland \&
  Shredchenko]{egorov2011three}
{\sc \au{Egorov, I.}, \au{Neiland, V.} \& \au{Shredchenko, V.}} \yr{2011}
  \at{Three-dimensional flow structures at supersonic flow over the compression
  ramp}.  \jt{AIAA 2011-730} .

\bibitem[Fletcher {\em et~al.\/}(2004)Fletcher, Ruban \&
  Walker]{fletcher_ruban_walker_2004}
{\sc \au{Fletcher, A. J.~P.}, \au{Ruban, A.~I.} \& \au{Walker, J. D.~A.}}
  \yr{2004}  \at{Instabilities in supersonic compression ramp flow}.
  \jt{Journal of Fluid Mechanics}  \bvol{517},  \pg{309–330}.

\bibitem[Gai \& Khraibut(2019)]{gai_khraibut_2019}
{\sc \au{Gai, S.~L.} \& \au{Khraibut, A.}} \yr{2019}  \at{Hypersonic
  compression corner flow with large separated regions}.  \jt{Journal of Fluid
  Mechanics}  \bvol{877},  \pg{471–494}.

\bibitem[Gaitonde(2015)]{GAITONDE201580}
{\sc \au{Gaitonde, D.~V.}} \yr{2015}  \at{Progress in shock wave/boundary layer
  interactions}.  \jt{Progress in Aerospace Sciences}  \bvol{72},  \pg{80--99},
  celebrating 60 Years of the Air Force Office of Scientific Research (AFOSR):
  A Review of Hypersonic Aerothermodynamics.

\bibitem[Ganapathisubramani {\em et~al.\/}(2009)Ganapathisubramani, Clemens \&
  Dolling]{GANAPATHISUBRAMANI_CLEMENS_DOLLING_2009}
{\sc \au{Ganapathisubramani, B.}, \au{Clemens, N.~T.} \& \au{Dolling, D.~S.}}
  \yr{2009}  \at{Low-frequency dynamics of shock-induced separation in a
  compression ramp interaction}.  \jt{Journal of Fluid Mechanics}  \bvol{636},
  \pg{397–425}.

\bibitem[Ginoux(1960)]{ginoux1960existence}
{\sc \au{Ginoux, J.~J.}} \yr{1960}  \bt{The existence of three-dimensional
  perturbations in the reattachment of a two-dimensional supersonic
  boundary-layer after separation}. {\em Tech. Rep.\/} TM 3.  \org{Von Karman
  Institute for Fluid Dynamics, Rhode-Saint-Genese (Belgium)}.

\bibitem[Ginoux(1966)]{ginoux1966laminar}
{\sc \au{Ginoux, J.~J.}} \yr{1966}  \bt{Laminar separation in supersonic and
  hypersonic flows}. {\em Tech. Rep.\/} AF EOAR 66-6.  \org{Training center for
  experimental aerodynamics Rhode-Saint-Genese (Belgium)}.

\bibitem[Ginoux(1969)]{ginoux1969some}
{\sc \au{Ginoux, J.~J.}} \yr{1969}  \bt{On some properties of reattaching
  laminar and transitional high speed flows.} {\em Tech. Rep.\/} VKI TN 53.
  \org{Von Karman Institute for Fluid Dynamics, Rhode-Saint-Genese (Belgium)}.

\bibitem[Grisham {\em et~al.\/}(2018)Grisham, Dennis \& Lu]{GrishamEtAl}
{\sc \au{Grisham, J.~R.}, \au{Dennis, B.~H.} \& \au{Lu, F.~K.}} \yr{2018}
  \at{Incipient separation in laminar ramp-induced shock-wave/boundary-layer
  interactions}.  \jt{AIAA Journal}  \bvol{56}~(2),  \pg{524--531}.

\bibitem[Guo {\em et~al.\/}(1995)Guo, Adams \& Kleiser]{GuoAdamsKleiser1995}
{\sc \au{Guo, Y.}, \au{Adams, N.~A.} \& \au{Kleiser, L.}} \yr{1995}
  \at{Modeling of nonparallel effects in temporal direct numerical simulations
  of compressible boundary-layer transition}.  \jt{Theoretical and
  Computational Fluid Dynamics}  \bvol{7}~(2),  \pg{141--157}.

\bibitem[Hao {\em et~al.\/}(2023)Hao, Cao, Guo \& Wen]{Hao_Cao_Guo_Wen_2023}
{\sc \au{Hao, J.}, \au{Cao, S.}, \au{Guo, P.} \& \au{Wen, C.-Y.}} \yr{2023}
  \at{Response of hypersonic compression corner flow to upstream disturbances}.
   \jt{Journal of Fluid Mechanics}  \bvol{964},  \pg{A25}.

\bibitem[Hao {\em et~al.\/}(2021)Hao, Cao, Wen \& Olivier]{hao_occurancer_2021}
{\sc \au{Hao, J.}, \au{Cao, S.}, \au{Wen, C.-Y.} \& \au{Olivier, H.}} \yr{2021}
   \at{Occurrence of global instability in hypersonic compression corner flow}.
   \jt{Journal of Fluid Mechanics}  \bvol{919},  \pg{A4}.

\bibitem[Holden(1966)]{Holden1966}
{\sc \au{Holden, M.~S.}} \yr{1966}  \at{Experimental studies of separated flows
  at hypersonic speeds. {II} - two-dimensional wedge separated flow studies.}
  \jt{AIAA Journal}  \bvol{4}~(5),  \pg{790--799}.

\bibitem[Holden \& Moselle(1969)]{Holden1970THEORETICALAE}
{\sc \au{Holden, M.~S.} \& \au{Moselle, J.}} \yr{1969}  \bt{Theoretical and
  experimental studies of the shock-wave boundary layer interaction on
  compression surfaces in hypersonic flow}.  \org{Cornell aeronautical
  laboratory report no:cal-af-2410-a-1, arl-7--0002}.

\bibitem[Hozumi {\em et~al.\/}(2001)Hozumi, Yamamoto, Fujii, Ledy, Devezeaux \&
  Fontaine]{Hozumi2001AIAA}
{\sc \au{Hozumi, K.}, \au{Yamamoto, Y.}, \au{Fujii, K.}, \au{Ledy, J.-P.},
  \au{Devezeaux, D.} \& \au{Fontaine, J.}} \yr{2001}  \at{Investigation of
  hypersonic compression ramp heating at high angles of attack}.  \jt{Journal
  of Spacecraft and Rockets}  \bvol{38}~(4),  \pg{488--496}.

\bibitem[Inger(2008)]{Inger2008}
{\sc \au{Inger, G.}} \yr{2008}  \at{Shock/boundary layer interaction in
  rarefied flow}.  \jt{AIAA Paper 2008-4168} .

\bibitem[Inger(2007)]{INGER200742}
{\sc \au{Inger, G.~R.}} \yr{2007}  \at{Triple-deck theory of supersonic laminar
  viscous–inviscid interaction due to wall temperature jumps}.  \jt{Progress
  in Aerospace Sciences}  \bvol{43}~(1),  \pg{42--63}.

\bibitem[Ivanov \& Rogasinsky(1988)]{majorant}
{\sc \au{Ivanov, M.~S.} \& \au{Rogasinsky, S.~V.}} \yr{1988}  \at{Analysis of
  the numerical techniques of the direct simulation {M}onte {C}arlo method in
  the rarefied gas dynamics}.  \jt{Soviet Journal of Numerical Analysis and
  Mathematical Modeling}  \bvol{3}~(6),  \pg{453--465}.

\bibitem[Karpuzcu {\em et~al.\/}(2022)Karpuzcu, Levin, Cerulus \&
  Theofilis]{KarpuzcuCRAviation2022}
{\sc \au{Karpuzcu, I.~T.}, \au{Levin, D.~A.}, \au{Cerulus, N.} \&
  \au{Theofilis, V.}} \yr{2022}  \at{On linear stability of compression corner
  flows obtained by kinetic theory}.  \jt{AIAA Paper 2022-4102} .

\bibitem[Karpuzcu {\em et~al.\/}(2023)Karpuzcu, Levin, Cerulus \&
  Theofilis]{KarpuzcuCRSciTech2023}
{\sc \au{Karpuzcu, I.~T.}, \au{Levin, D.~A.}, \au{Cerulus, N.} \&
  \au{Theofilis, V.}} \yr{2023}  \at{On the unsteadiness and three
  dimensionality of a laminar separation bubble for a supersonic flow over a
  compression corner}.  \jt{AIAA Paper 2023-0679} .

\bibitem[Klineberg(1968)]{KlinebergPhDThesis}
{\sc \au{Klineberg, J.~M.}} \yr{1968}  \at{Theory of laminar viscous-inviscid
  interactions in supersonic flow}. PhD thesis, California Institute of
  Technology.

\bibitem[Klineberg \& Lees(1969)]{KlinebergAIAAJ}
{\sc \au{Klineberg, J.~M.} \& \au{Lees, L.}} \yr{1969}  \at{Theory of laminar
  viscous-inviscid interactions in supersonic flow}.  \jt{AIAA Journal}
  \bvol{7}~(12),  \pg{2211--2221}.

\bibitem[Mack(1969)]{Mack1969}
{\sc \au{Mack, L.~M.}} \yr{1969}  \bt{Boundary-layer stability theory}.
  \org{Jet propulsion laboratory, document no. 900–277, rev. a}.

\bibitem[Mack(1984)]{Mack1984}
{\sc \au{Mack, L.~M.}} \yr{1984}  \bt{Boundary-layer linear stability theory}.
  \org{Agard report no. 709, part 3}.

\bibitem[Mallinson {\em et~al.\/}(1997)Mallinson, Gai \&
  Mudford]{mallinson_gai_mudford_1997}
{\sc \au{Mallinson, S.~G.}, \au{Gai, S.~L.} \& \au{Mudford, N.~R.}} \yr{1997}
  \at{The interaction of a shock wave with a laminar boundary layer at a
  compression corner in high-enthalpy flows including real gas effects}.
  \jt{Journal of Fluid Mechanics}  \bvol{342},  \pg{1–35}.

\bibitem[Mayer {\em et~al.\/}(2011)Mayer, Wernz \&
  Fasel]{MAYER_WERNZ_FASEL_2011}
{\sc \au{Mayer, C.}, \au{Wernz, S.} \& \au{Fasel, H.~F.}} \yr{2011}
  \at{Numerical investigation of the nonlinear transition regime in a {M}ach 2
  boundary layer}.  \jt{Journal of Fluid Mechanics}  \bvol{668},
  \pg{113–149}.

\bibitem[Messiter(1970)]{Messiter}
{\sc \au{Messiter, A.~F.}} \yr{1970}  \at{Boundary-layer flow near the trailing
  edge of a flat plate}.  \jt{SIAM Journal on Applied Mathematics}
  \bvol{18}~(1),  \pg{241--257}.

\bibitem[Neiland(1969)]{neiland1969theory}
{\sc \au{Neiland, V.~Y.}} \yr{1969}  \at{Theory of laminar boundary layer
  separation in supersonic flow}.  \jt{Fluid Dynamics}  \bvol{4}~(4),
  \pg{33--35}.

\bibitem[Neiland {\em et~al.\/}(2004)Neiland, Bogolepov, Dudin \&
  Lipatov]{NeilandBogolepovDudinLipatov}
{\sc \au{Neiland, V.~Y.}, \au{Bogolepov, V.}, \au{Dudin, G.} \& \au{Lipatov,
  I.}} \yr{2004} {\em Asymptotic Theory of Viscous Supersonic Gas Flows\/}.
  \publ{Fizmatlit Moscow}.

\bibitem[Pagella {\em et~al.\/}(2004)Pagella, Babucke \& Rist]{pagella2004}
{\sc \au{Pagella, A.}, \au{Babucke, A.} \& \au{Rist, U.}} \yr{2004}
  \at{Two-dimensional numerical investigations of small-amplitude disturbances
  in a boundary layer at {Ma}= 4.8: Compression corner versus impinging shock
  wave}.  \jt{Physics of Fluids}  \bvol{16}~(7),  \pg{2272--2281}.

\bibitem[Paredes {\em et~al.\/}(2016)Paredes, Gosse, Theofilis \&
  Kimmel]{paredes_gosse_theofilis_kimmel_2016}
{\sc \au{Paredes, P.}, \au{Gosse, R.}, \au{Theofilis, V.} \& \au{Kimmel, R.}}
  \yr{2016}  \at{Linear modal instabilities of hypersonic flow over an elliptic
  cone}.  \jt{Journal of Fluid Mechanics}  \bvol{804},  \pg{442–466}.

\bibitem[Quintanilha {\em et~al.\/}(2022)Quintanilha, Paredes, Hanifi \&
  Theofilis]{quintanilha2022transient}
{\sc \au{Quintanilha, H.}, \au{Paredes, P.}, \au{Hanifi, A.} \& \au{Theofilis,
  V.}} \yr{2022}  \at{Transient growth analysis of hypersonic flow over an
  elliptic cone}.  \jt{Journal of Fluid Mechanics}  \bvol{935},  \pg{A40}.

\bibitem[Ringuette {\em et~al.\/}(2009)Ringuette, Bookey, Wyckham \&
  Smits]{Ringuette2009AIAAJ}
{\sc \au{Ringuette, M.~J.}, \au{Bookey, P.}, \au{Wyckham, C.} \& \au{Smits,
  A.~J.}} \yr{2009}  \at{Experimental study of a {M}ach 3 compression ramp
  interaction at ${R}e_{\theta}$ = 2400}.  \jt{AIAA Journal}  \bvol{47}~(2),
  \pg{373--385}.

\bibitem[Rist \& Fasel(1995)]{Rist_Fasel_1995}
{\sc \au{Rist, U.} \& \au{Fasel, H.}} \yr{1995}  \at{Direct numerical
  simulation of controlled transition in a flat-plate boundary layer}.
  \jt{Journal of Fluid Mechanics}  \bvol{298},  \pg{211–248}.

\bibitem[Rizzetta {\em et~al.\/}(1978)Rizzetta, Burggraf \&
  Jenson]{rizzetta1978triple}
{\sc \au{Rizzetta, D.}, \au{Burggraf, O.} \& \au{Jenson, R.}} \yr{1978}
  \at{Triple-deck solutions for viscous supersonic and hypersonic flow past
  corners}.  \jt{Journal of Fluid Mechanics}  \bvol{89}~(3),  \pg{535--552}.

\bibitem[Robinet(2007)]{robinet_2007}
{\sc \au{Robinet, J.-C.}} \yr{2007}  \at{Bifurcations in
  shock-wave/laminar-boundary-layer interaction: global instability approach}.
  \jt{Journal of Fluid Mechanics}  \bvol{579},  \pg{85–112}.

\bibitem[Roghelia {\em et~al.\/}(2017)Roghelia, Olivier, Egorov \&
  Chuvakhov]{Roghelia2017}
{\sc \au{Roghelia, A.}, \au{Olivier, H.}, \au{Egorov, I.} \& \au{Chuvakhov,
  P.}} \yr{2017}  \at{Experimental investigation of {G}örtler vortices in
  hypersonic ramp flows}.  \jt{Experiments in Fluids}  \bvol{58}~(10),
  \pg{139}.

\bibitem[Sawant {\em et~al.\/}(2021)Sawant, Levin \& Theofilis]{sawant_POF}
{\sc \au{Sawant, S.~S.}, \au{Levin, D.~A.} \& \au{Theofilis, V.}} \yr{2021}
  \at{A kinetic approach to studying low-frequency molecular fluctuations in a
  one-dimensional shock}.  \jt{Physics of Fluids}  \bvol{33}~(10),
  \pg{104106}.

\bibitem[Sawant {\em et~al.\/}(2022{\natexlab{{\em a\/}}})Sawant, Levin \&
  Theofilis]{sawant_TCFD}
{\sc \au{Sawant, S.~S.}, \au{Levin, D.~A.} \& \au{Theofilis, V.}}
  \yr{2022{\natexlab{{\em a\/}}}}  \at{{Analytical prediction of low-frequency
  fluctuations inside a one-dimensional shock}}.  \jt{Theoretical and
  Computational Fluid Dynamics}  \bvol{36}~(1),  \pg{25--40}.

\bibitem[Sawant {\em et~al.\/}(2022{\natexlab{{\em b\/}}})Sawant, Theofilis \&
  Levin]{sawant_etal_2022}
{\sc \au{Sawant, S.~S.}, \au{Theofilis, V.} \& \au{Levin, D.}}
  \yr{2022{\natexlab{{\em b\/}}}}  \at{On the synchronisation of
  three-dimensional shock layer and laminar separation bubble instabilities in
  hypersonic flow over a double wedge}.  \jt{Journal of Fluid Mechanics}
  \bvol{941},  \pg{A7}.

\bibitem[Sawant {\em et~al.\/}(2018)Sawant, Tumuklu, Jambunathan \&
  Levin]{Sawant2018}
{\sc \au{Sawant, S.~S.}, \au{Tumuklu, O.}, \au{Jambunathan, R.} \& \au{Levin,
  D.~A.}} \yr{2018}  \at{Application of adaptively refined unstructured grids
  in dsmc to shock wave simulations}.  \jt{Computers and Fluids}  \bvol{170},
  \pg{197--212}.

\bibitem[Seddougui \& Bassom(1997)]{SeddouguiBassom1997}
{\sc \au{Seddougui, S.~O.} \& \au{Bassom, A.~P.}} \yr{1997}  \at{Instability of
  hypersonic flow over a cone}.  \jt{Journal of Fluid Mechanics}  \bvol{345},
  \pg{383–411}.

\bibitem[Settles \& Dodson(1994)]{Settles1994AIAAJ}
{\sc \au{Settles, G.~S.} \& \au{Dodson, L.~J.}} \yr{1994}  \at{Supersonic and
  hypersonic shock/boundary-layer interaction database}.  \jt{AIAA Journal}
  \bvol{32}~(7),  \pg{1377--1383}.

\bibitem[Settles {\em et~al.\/}(1979)Settles, Fitzpatrick \&
  Bogdonoff]{Settles1979AIAAJ}
{\sc \au{Settles, G.~S.}, \au{Fitzpatrick, T.~J.} \& \au{Bogdonoff, S.~M.}}
  \yr{1979}  \at{Detailed study of attached and separated compression corner
  flowfields in high {R}eynolds number supersonic flow}.  \jt{AIAA Journal}
  \bvol{17}~(6),  \pg{579--585}.

\bibitem[Simeonides(1992)]{SimeonidesPhDThesis}
{\sc \au{Simeonides, G.}} \yr{1992}  \at{Hypersonic shock wave boundary layer
  interactions over compression corners}. PhD thesis, University of Bristol and
  Von Karman Institute for Fluid Dynamics.

\bibitem[Simeonides \& Haase(1995)]{Simeonides_Haase_1995}
{\sc \au{Simeonides, G.} \& \au{Haase, W.}} \yr{1995}  \at{Experimental and
  computational investigations of hypersonic flow about compression ramps}.
  \jt{Journal of Fluid Mechanics}  \bvol{283},  \pg{17–42}.

\bibitem[Simeonides {\em et~al.\/}(1994)Simeonides, Haase \&
  Manna]{SimeonidesetalAIAAJ1994}
{\sc \au{Simeonides, G.}, \au{Haase, W.} \& \au{Manna, M.}} \yr{1994}
  \at{Experimental, analytical, and computational methods applied to hypersonic
  compression ramp flows}.  \jt{AIAA Journal}  \bvol{32}~(2),  \pg{301--310}.

\bibitem[Stewartson(1970)]{stewartson1970laminar}
{\sc \au{Stewartson, K.}} \yr{1970}  \at{On laminar boundary layers near
  corners}.  \jt{The Quarterly Journal of Mechanics and Applied Mathematics}
  \bvol{23}~(2),  \pg{137--152}.

\bibitem[Theofilis(2003)]{Theofilis2003}
{\sc \au{Theofilis, V.}} \yr{2003}  \at{Advances in global linear instability
  analysis of nonparallel and three-dimensional flows}.  \jt{Progress in
  Aerospace Sciences}  \bvol{39},  \pg{249--315}.

\bibitem[Theofilis(2011)]{theofilisARFM}
{\sc \au{Theofilis, V.}} \yr{2011}  \at{Global linear instability}.  \jt{Annual
  Review of Fluid Mechanics}  \bvol{43}~(1),  \pg{319--352}.

\bibitem[Theofilis(2020)]{theofilis2020massively}
{\sc \au{Theofilis, V.}} \yr{2020} Massively parallel solution of the global
  linear instability nonsymmetric complex generalized eigenvalue problem.
  \bt{In {\em 60th Israel Annual Conference on Aerospace Sciences, IACAS\/}}, ,
   \vol{vol.~60},  \pg{pp. 801--806}.

\bibitem[Theofilis {\em et~al.\/}(2000)Theofilis, Hein \&
  Dallmann]{Theofilis2000}
{\sc \au{Theofilis, V.}, \au{Hein, S.} \& \au{Dallmann, U.}} \yr{2000}  \at{On
  the origins of unsteadiness and three-dimensionality in a laminar separation
  bubble}.  \jt{Philosophical Transactions of the Royal Society of London}
  \bvol{358},  \pg{3229--3324}.

\bibitem[Tumuklu {\em et~al.\/}(2019)Tumuklu, Levin \&
  Theofilis]{TumukluPRF2019}
{\sc \au{Tumuklu, O.}, \au{Levin, D.~A.} \& \au{Theofilis, V.}} \yr{2019}
  \at{Modal analysis with proper orthogonal decomposition of hypersonic
  separated flows over a double wedge}.  \jt{Phys. Rev. Fluids}  \bvol{4},
  \pg{033403}.

\bibitem[Tumuklu {\em et~al.\/}(2018)Tumuklu, Theofilis \& Levin]{tumukluPoF2}
{\sc \au{Tumuklu, O.}, \au{Theofilis, V.} \& \au{Levin, D.~A.}} \yr{2018}
  \at{On the unsteadiness of shock laminar boundary layer interactions of
  hypersonic flows over a double cone}.  \jt{Physics of Fluids}
  \bvol{30}~(10),  \pg{106111}.

\end{thebibliography}
\bibliographystyle{jfm}
\end{document}